\documentclass[a4paper,11pt]{article}
\usepackage[latin1]{inputenc}
\usepackage[T1]{fontenc}
\usepackage[english]{babel}
\usepackage{graphicx}
\usepackage{epsf}
\begin{document}
\title{Properties of homogeneous cosmologies in scalar tensor theories} 
\author{Stéphane Fay\\
Laboratoire Univers et Théories (LUTH), CNRS-UMR 8102\\
Observatoire de Paris, F-92195 Meudon Cedex\\
FRANCE\\
\small{Steph.Fay@Wanadoo.fr}} 
\maketitle
\begin{center}
\begin{abstract} 
We study the isotropisation of the homogeneous but anisotropic Bianchi class $A$ models in presence of a minimally coupled and massive scalar field with or without a perfect fluid. To this end, we use the Hamiltonian formalism of Arnowitt, Deser and Misner(ADM) and the dynamical systems analysis methods. Our results allow to define three kinds of isotropisation called class 1, 2 and 3. We have specifically studied the class 1 and obtained some general constraints on scalar-tensor theories which are necessary conditions for isotropisation. The asymptotical behaviors of the metric functions and potential in the neighborhood of isotropy have also been determined when the isotropic state is reached sufficiently quickly. We show that the scalar field responsible for isotropisation may be quintessent and that the presence of curvature favor a late times acceleration and quintessence. Some applications are made with the well known exponential law potential by using our theoretical results but also by help of numerical analysis. The isotropisation process with a power law potential is also explored. We think this work represents a framework able to guide some future researches on the isotropisation of homogeneous models in scalar-tensor theories and we argue by discussing briefly about some recent results we have obtained in presence of a non minimally coupled scalar field or several scalar fields.\end{abstract} 
\end{center}
\emph{Please, review the English for style}
%--------------------------------------------------------------------------------------------------------------------------------------------------------------------%
\section{Introduction} 
Two basic elements used to define a cosmology are a geometry represented mathematically by a metric and a material contents described mathematically by a Lagrangian. In this work, we will consider the isotropisation of the homogeneous but anisotropic cosmologies of Bianchi in presence of a scalar field. Let us begin by explaining what a scalar field is and why it seems impossible to
circumvent in cosmology.\\\\
By definition, a scalar field is a function which at each point of space and time associates a
number. A good example is the temperature of a room: in each one of its points, one can associate a number defining a temperature. Another example is the gravitational potential $\phi$ outside a mass
$M$. These fields abound in particles physics and although they were not detected yet, one can advance some arguments making their prediction natural. Hence, the best experimentally established theories rest on vectors and tensors fields: 
\begin{itemize} 
\item the electromagnetic interaction is described by a neutral and massless vectors field corresponding to a photon without any charge and an interaction of infinite range.
\item the weak interaction is described by a massive vectors field representing the W and Z bosons. It means that it is with short range and unstable: these particles disintegrate in pairs of other particles. 
\item the strong interaction is described by a massive vectors field of gluons. 
\item the gravitational interaction is described by a tensors field corresponding to graviton.\end{itemize} 
All these vectors and tensors fields correspond to particles with integer spin being respectively $1\hbar$\footnote{$\hbar$ is the constant of Planck divided by $2\pi$} and $2\hbar$, called bosons\footnote{Particles with half spin belongs to the family of the fermions.}. It then seems natural to wonder what a boson of spin $0$ would represent: the answer is "a scalar field".\\ 
Probably the most known is the field of Higgs-Englert, introduced in order to solve a renormalisation problem\cite{Zel86} involved by the weak force: rather than to impose a mass on bosons propagating the weak force, it is supposed that the vectors field representing them interacts with a massive scalar field $\phi$, i.e. having a potential which describes the interaction with the vectors field. The form of this potential would be $U(\phi)=k(\phi^2-\phi_0^2)$. This field, which one hopes for detection by the LHC in 2008, gives their mass to the particles which interact with it.\\
Reasons different from those of Higgs can be evoked concerning the presence of scalar fields in particles physics. Thus, the supersymmetric theories postulate the equality between the fermionic and bosonic degrees of freedom: to each boson(of which that of Higgs), a fermion corresponds and vice versa. However, this can be done only by adding some degrees of freedom which are introduced with help of scalar fields whose potentials $U(\phi)$ can be completely different from the one above. Generally these scalar fields are called dilaton. The supersymmetric particles could be essential components of the dark matter. In particular the lightest of the neutralinos, a state resulting from a mixture of several supersymmetric particles, could be the lightest of the supersymmetric particles and a candidate for cold dark matter. For a popular introduction to the supersymmetry theory, one will refer to the book of Gordon Kane\cite{Kan00}, "Supersymmetry, unveiling the ultimate laws of Nature".\\
If these fields are present in particles physics, it is likely that they populate the Universe and thus must, like any material contents, have some cosmological consequences. The first relativistic scalar-tensor theory was the Jordan, Brans and Dicke theory\cite{BraDic61} in the sixties which was issued\cite{Bra97} of the ideas of Kaluza-Klein\cite{Kal21}, Dirac\cite{Dir37} and Mach. More modern reasons to consider scalar fields in cosmology appear at the beginning of the eighties: the ideas of Guth on inflation gave birth to scalar fields called inflatons. At that time cosmology suffers from many conceptual problems: why the Universe seems so flat? How causally disconnected areas at the beginning of times can be so similar today? Guth\cite{Gut81} noticed that they would be partially solved if, at early times, the Universe undergoes a period of inflation with an exponential expansion. For that, the first idea which comes to mind is to introduce a cosmological constant. However the observations show that its current value would be $10^{120}$ time smaller than that predicted at early times: it is the cosmological constant problem\cite{Car03}. A way to solve it is to consider a massive scalar field, i.e. having a potential representing the coupling of the field with itself and able to play the role of a variable cosmological constant. Since the end of the nineties, the presence of this potential found new reasons to exist with detection by two independent teams\cite{Per99, Rie98} of the Universe expansion acceleration. One of the most widespread explanations of this phenomenon would be the presence of a  quintessent scalar field whose density and pressure are bound by an equation of state similar to that of a perfect fluid and whose barotropic index would be negative. It then results a negative scalar field pressure which would be at the origin of this new and recent period of expansion acceleration.\\\\
With regard to the geometrical description of the Universe, we will be interested by the 
homogeneous but anisotropic cosmological models of Bianchi for which the expansion rate depends on the observation direction: it is thus an anisotropic Universe on the contrary traditional models
of Friedmann-Lemaître-Robertson-Walker (FLRW) where the expansion is the same whatever the observation direction. Let us start with a short history of these models. Luigi Bianchi\footnote{These information come from http://www34.homepage.villanova.edu/robert.jantzen/ bianchi/bianchi.html\#papers.} was born in Parma on June 18, 1856. It was the student of Ulisse Dini and Enrico Batti at the Ecole Normale of Pisa
and became professor at the University of Pisa in 1886 then the Ecole Normale director in 1918 until its death in 1928. Its mathematical contributions were published in 11 volumes by the Italian Mathematical Union and cover a great number of fields. With regard to the riemanian geometry, he is especially known for its discovery of the Bianchi identities whose full demonstration was given in
\cite{Bia02}(it had discovered them first once in an article of 1888 but had neglected their importance by giving them in footnote). In 1897, by using the results of Lipshitz\cite{Lip70} and
Killing\cite{Kil92} as well as the Lie\cite{LieEng88} theory of continuous groups, he gave a complete classification of the Riemann 3-varieties isometries classes, identified by the roman letters $I$ to $IX$. At the time, neither general relativity, nor special relativity existed yet\footnote{It also exists the axisymmetric anisotropic model of Kantowski and Sachs}. In 1951, the work of Bianchi was introduced in cosmology by Abraham Taub into his article "Empty Spacetimes Admitting has Three-Parameter Group of Motions"\cite{Tau51}. Spatially homogeneous spacetime  have a time dependant space geometry which is thus a homogeneous 3-geometry. Thus, spacetime has a group of isometry with $r$ dimensions acting on a family of hypersurface with $r=3$(simply transitive action), $r=4$(locally rotationally symmetric) or $r=6$(isotropic models). The case $r=3$ became known under the name of Bianchi cosmologies after the article of Taub. During one decade, the Bianchi models felt into forgets until the rebirth of General Relativity at the beginning of the sixties. O. Heckmann and E. Schücking made them appear again in 1958 in their work "Gravitation,
an Introduction to Current Research"\cite{HecSch62}. Then it was with the turn of the Russian school of Lifshitz and Khalatnikov, joined later by Belinsky (BKL), through their study on the chaotic approach of the initial singularity which inspired Misner in the USA and later Hawking and Ellis in the United Kingdom. The Bianchi classification was reexamined by C.G. Behr in a work not published but reported in \cite{EstWahBeh68} in 1968. Finally an essential contribution on the Bianchi models was made by Ryan, a student of Misner, and summarized through its book "Homogenous Relativistic Cosmologies"\cite{RyaShe75}.\\
Let us explain the physical reasons which, in our opinion, justify the study of these models. The Universe such as we observe it today is very well described by the isotropic and homogeneous cosmological models of Friedman-Lemaitre-Robertson-Walker (FLRW). It was shown by the observations of the cosmic microwave background (CMB) by the satellites COBE and WMAP\cite{Spe03}. However nothing enables us to extrapolate these properties of isotropy and homogeneity at early times before the radiation/matter decoupling. The question thus arises of knowing why the Universe has these properties whereas there is an infinity of cosmological models not having them. Primarily, two answers are distinguished: 
\begin{itemize} 
\item It could exist a quantum principle which selects, among the set of the possible
models, the FLRW ones as being the most probable. This answer rests on the development of a quantum theory of the initial conditions. 
\item the early Universe would be inhomogeneous and anisotropic but its evolution would lead it to a homogeneous and isotropic state (asymptotic or temporary) corresponding to FLRW models.
\end{itemize}
It is this second point of view which we will adopt by considering that the Universe is initially anisotropic and asymptotically becomes isotropic. We will keep the assumption of homogeneity because the inhomogeneous cosmological models are not classified due to their lack of symmetry. Moreover we will consider that this state is reached asymptotically and not transitory. Indeed, the observations show that our Universe must be isotropic since at least its first million years and
it is thus reasonable to suppose that once this state reached, it is stable. To consider that the Universe is initially anisotropic thus makes it possible to explain the isotropisation driving process rather than to consider this state in an ad hoc way. Another advantage of the anisotropic models resides in their approach of the singularity in General relativity which is oscillatory and chaotic for the most general of them. It would be shared, according to the BKL conjecture, by the inhomogeneous and anisotropic models\cite{UggElsWaiEll03} on the contrary to the FRLW models whose singularity approach is monotonous.\\\\ 
In this chapter, we will be interested by the isotropisation process for the Bianchi class $A$ models with the aim to constrain the scalar-tensor theories so that they led the Universe to a stable isotropic state. For that we will use a method inspired from Wainwright, Ellis et al book "Dynamical Systems in Cosmology"\cite{WaiEll97}. In this book, the dynamical evolution of homogeneous cosmologies without scalar field is described using first order equations systems obtained thanks to the orthonormal formalism. These equations are then studied from the point of view of their equilibrium points. We will proceed in the same way but we will use the Arnowitt-Deser-Misner(ADM) Hamiltonian formalism and will seek the equilibrium points representing a stable isotropic state for the Universe.\\
Most of the results which will follow were published in \cite{Fay01, Fay01A, Fay03}. This chapter is thus for us an occasion to gather them and to improve them considerably. Indeed, since these first papers on isotropisation, some significant progress which we will report here, were made. They consist of: 
\begin{itemize} 
\item a classification of the Universe stable isotropic states in three classes. 
\item a clarification of our results range thanks to a discussion on their stability. 
\item the use of numerical methods allowing to validate our calculations and to represent them graphically. 
\item some new physical interpretations of our calculations and results. 
\end{itemize} 
The plan of this work is as follows. In the section \ref{s0}, we mathematically describe the
geometrical and physical frameworks of this work by writing the metric of the Bianchi models, the action and ADM Hamiltonian of the scalar-tensor theories. In the sections \ref{s2} and \ref{s3}, we study the isotropisation of the Bianchi class $A$ models. In the section \ref{s4}, we analyse the  quintessent asymptotical nature of the scalar field. We conclude in the section \ref{s5}.
%--------------------------------------------------------------------------------------------------------------------------------------------------------------------%
\section{Background}\label{s0} 
This section mathematically describes the geometrical and physical framework of this work, namely the Bianchi models metric, the action and ADM Hamiltonian of the scalar-tensor theories and the method which we will systematically use to study the isotropisation process.
%--------------------------------------------------------------------------------------------------------------------------------------------------------------------%
\subsection{Metric} 
The metric that we will use reflects the 3+1 decomposition of spacetime, necessary to obtain the ADM Hamiltonian. It is written: 
\begin{equation}\label{metric} 
ds^2 = -(N^2 - N_iN^i)d\Omega^2 + 2N_i d\Omega\omega^i + R_0 ^2 g_{ij}\omega^i \omega^j
\end{equation}
$N$ being the lapse function, $N_i$ the shift functions and $\omega^i$ the 1-forms generating the Bianchi homogeneous spaces. For example for the spatially flat Bianchi type $I$ model, we have $\omega^1=dx$, $\omega^2=dy$ and $\omega^3=dz$. We will choose a diagonal metric such as $N_i=0$ and the relation between the proper time $t$ and the variable $\Omega$ will be then: 
%-equation-%
\begin{equation}\label{time} 
dt=-Nd\Omega 
\end{equation}
%--------------------------------------------------------------------------------------------------------------------------------------------------------------------%
\subsection{Action} 
The general form of the action of a minimally coupled and massive scalar-tensor theory with a perfect fluid is written: 
%-equation-%
\begin{eqnarray} \label{action} 
S=(16\pi)^{-1}\int \mbox{[}R-(3/2+\omega)\phi^{, \mu}\phi_{,\mu}\phi^{-2}-U\mbox{]}\sqrt{-g}d^4 x+S_m(g_{ij})
\end{eqnarray} 
where the following functions are defined:
\begin{itemize}
\item $R$ is the curvature scalar. 
\item $\phi$ is the scalar field. Here, it is minimally coupled because there is no coupling between $\phi$ and the curvature and thus the gravitation constant does not vary with respect to this field. 
\item $\omega(\phi)$ is the Brans-Dicke function describing the coupling of the scalar field
with the metric.
\item $U$ is the potential describing the coupling of the scalar field with itself. When the potential is not zero, the scalar field is known as massive otherwise it is massless.
\item  $S_m(g_{ij})$ is the action representing the presence of a non tilted perfect fluid with an
equation of state $p=(\gamma-1)\rho$, where $p$ is its pressure, $\rho$ its density and $\gamma$ its barotropic index such as $\gamma\in\left[1,2\right]$. The case $\gamma=1$ represents a dust fluid and $\gamma=4/3$, a radiative fluid. Note that $S_m$ does not depend on $\phi$. If it were the case, the matter will not follow the spacetime geodesics and it is likely that such a deviation would have been observed.
\\
\end{itemize}
The conservation of the perfect fluid energy-momentum tensor allows to get: 
$$
\rho=V^{-\gamma}
$$
where $V=e^{-3\Omega}$ is the Universe 3-volume. One will also define the pressure and the density of the
scalar field like:
%-equation-%
\begin{eqnarray}\label{density}
\rho_\phi=\frac{1}{2}\frac{3/2+\omega}{\phi^2}\phi'^2+\frac{1}{2}U
\end{eqnarray} 
%-equation-% 
\begin{eqnarray}\label{pressure}
p_\phi=\frac{1}{2}\frac{3/2+\omega}{\phi^2}\phi'^2-\frac{1}{2}U
\end{eqnarray} 
where a prime indicates a derivative with respect to the proper time $t$.
%--------------------------------------------------------------------------------------------------------------------------------------------------------------------%
\subsection{Hamiltonian} 
In this subsection we look for the ADM Hamiltonian $H$ corresponding to the action (\ref{action}). We follow the procedure described in \cite{Nar72} and \cite{MatRyaTot73}. First, we rewrite the action as:
%-----------------------EQUATION-------------------------------------%
\begin{equation}\label{action2}
S=(16\pi)^{-1}\int(\Pi^{ij}\frac{\partial{g_{ij}}}{\partial{t}}+\Pi^{\phi}\frac{\partial{\phi}}{\partial{t}}-NC^0-N_iC^i)d^4x
\end{equation}
The $\Pi_{ij}$ and $\Pi_\phi$ are respectively the conjugate momenta of the metric functions and scalar field. The lapse and shift functions here play the role of Lagrange multipliers. By varying (\ref{action2}) with respect to $N$ and $N_i$, we get the constraints $C^0=0$ and $C^i=0$ with:
%-------------------------EQUATION-------------------------------------%
\begin{eqnarray}\label{C0}
C^0 &=&-\sqrt{^{(3)}g}^{(3)}R-\frac{1}{\sqrt{^{(3)}g}}(\frac{1}{2}(\Pi^k _k )^2 -\Pi^{ij}\Pi_{ij})+\frac{1}{\sqrt{^{(3)}g}}\frac{\Pi_\phi ^2 \phi^2 }{6+4\omega}+\nonumber\\
&&\sqrt{^{(3)}g}U(\phi)+\frac{1}{\sqrt{^{(3)}g}}\frac{\delta e^{3(\gamma-2)\Omega}}{24\pi^2}\\\nonumber
\end{eqnarray}
%----------------------------------EQUATION----------------------------------------%
\begin{equation}
C^i =\Pi^{ij}_{\mid j}
\end{equation}
where the "$^{(3)}$" stands for the quantities defined on an hypersurface and $\delta$ is a positive constant. We then rewrite the metric functions $g_{ij}$ as 
$$
e^{-2\Omega+2\beta_{ij}}
$$
It means that $\Omega$ stands for the isotropic part of the metric and $\beta_{ij}$ for the anisotropic parts. Then, using the Misner parameterization \cite{Mis62}:
%-----------------------EQUATION-------------------------------------%
\begin{equation}
p_k^i=2\pi\Pi_k^i-2/3\pi\delta_k^i\Pi_l^l
\end{equation}
%-----------------------EQUATION-------------------------------------%
\begin{equation}
6p_{ij}=diag(p_++\sqrt{3}p_-,p_+-\sqrt{3}p_-,-2p_+)
\end{equation}
%-----------------------EQUATION-------------------------------------%
\begin{equation}
\beta_{ij}=diag(\beta_++\sqrt{3}\beta_-,\beta_+-\sqrt{3}\beta_-,-2\beta_+)
\end{equation}
we rewrite the action under the form
$$
S=\int p_+d\beta_++p_-d\beta_-+p_\phi d\phi-Hd\Omega
$$
with $p_\phi=\pi\Pi_\phi$ and $H=2\pi\Pi_k^k$, the ADM Hamiltonian. Using the expression (\ref{C0}) and the constraint $C^0=0$, we finally get the form of $H$:
%----------------------------------EQUATION---------------------------%
\begin{equation}\label{hamiltonien}
H^2 = p_+ ^2 +p_- ^2 +12\frac{p_\phi ^2 \phi^2}{3+2\omega}+24\pi^2 R_0 ^6 e^{-6\Omega}U+V(\Omega,\beta_+,\beta_-)+\delta  e^{3(\gamma-2)\Omega}
\end{equation}
where $V(\Omega,\beta_+,\beta_-)$ stands for the curvature terms specifying each Bianchi model and  given in table \ref{tab1}.
\begin{table}[h]
\begin{center}
\begin{tabular}{|l|l|}
\hline
Bianchi type&$V(\Omega,\beta_+,\beta_-)$\\
\hline
$I$&0\\
\hline
$II$&$12\pi^2R_0^4e^{4(-\Omega+\beta_++\sqrt{3}\beta_-)}$\\
\hline
$VI_0$, $VII_0$&$24\pi^2R_0^4e^{-4\Omega+4\beta_+}(\cosh{4\sqrt{3}\beta_-}\pm 1)$\\
\hline
$VIII$, $IX$&$24\pi^2R_0^4e^{-4\Omega}\mbox{[}e^{4\beta_+}(\cosh{4\sqrt{3}\beta_-}-1)+$\\
&$1/2e^{-8\beta_+}\pm2e^{-2\beta_+}\cosh{2\sqrt{3}\beta_-}\mbox{]}$\\
\hline
\end{tabular}
\caption{\label{tab1}\scriptsize{Form of the curvature potential $V(\Omega,\beta_+,\beta_-)$ for each Bianchi model.}}
\end{center}
\end{table}
%--------------------------------------------------------------------------------------------------------------------------------------------------------------------%
\subsection{Method} 
The method that we will use to study the Bianchi models isotropisation process will be as follows:
\begin{itemize} 
\item From ADM Hamiltonian (\ref{hamiltonien}), we determine the Hamiltonian first order field equations. 
\item In order to use the dynamical systems methods\cite{WaiEll97}, we rewrite these equations using bounded variables. 
\item We then look for the equilibrium points representing the stable isotropic states. We deduce some constraints on the functions $\omega$ and $U$ defining the properties of the scalar field and we calculate the exact solutions corresponding to the equilibrium points. 
\item We apply our results to some scalar-tensor theories usually studied in the literature,
which will enable us to test their validity.
\end{itemize}
%--------------------------------------------------------------------------------------------------------------------------------------------------------------------%
\section{The flat Bianchi type $I$ model}\label{s2} 
The Bianchi type $I$ model is the only one to be spatially flat. It thus contains the flat FLRW solutions. In the first subsection, we will study the isotropisation process without a perfect fluid and then, in a second subsection, we will consider a perfect fluid. We proceeded so that these two subsections can be approached separately: in each of them we wrote the field equations and discussed our results.
%--------------------------------------------------------------------------------------------------------------------------------------------------------------------%
\subsection{Vacuum case}\label{s212}
%--------------------------------------------------------------------------------------------------------------------------------------------------------------------%
\subsubsection{Field equations}\label{s2121} 
The ADM Hamiltonian of the Bianchi type $I$ model when no perfect fluid is present but with a minimally coupled and massive scalar field is written: 
%-equation-%
\begin{equation} 
H^2 = p_+ ^2 +p _ - ^2 +12\frac{p_\phi ^2\phi^2}{3+2\omega}+24\pi^2 R_0 ^6 e^{-6\Omega}U 
\end{equation} 
It comes then for the hamiltonian equations: 
%-equation-%
\begin{equation} \dot{\beta} _ \pm = \frac{\partial H}{\partial p _
\pm}=\frac{p_\pm}{H} \end{equation} 
%-equation-% 
\begin{equation}
\dot{\phi}=\frac{\partial H}{\partial p_\phi}=\frac{12\phi^2
p_\phi}{(3+2\omega)H} \end{equation} 
%-equation-% 
\begin{equation}
\dot{p}_\pm=-\frac{\partial H}{\partial \beta _ \pm}=0 \end{equation}
%-equation-% 
\begin{equation} 
\dot{p}_\phi=-\frac{\partial H}{\partial \phi}=-12\frac{\phi p_\phi ^2}{(3+2\omega)H}+12\frac{\omega_\phi \phi^2 p_\phi ^2}{(3+2\omega)^2 H}-12\pi^2 R_0 ^6 \frac{e^{-6\Omega}U_\phi}{H} \end{equation} 
%-equation-%
\begin{equation} 
\dot{H}=\frac{dH}{d\Omega}=\frac{\partial H}{\partial\Omega}=-72\pi^2 R_0 ^6 \frac{e^{-6\Omega}U}{H} \end{equation} 
a dot meaning a derivative with respect to $\Omega$. Having chosen $N_i=0$ and using the fact that $\partial \sqrt{g}/\partial \Omega=-1/2\Pi^k_k N$\cite{Nar72}, one deduces the form of the lapse function:
%-equation-%
\begin{equation}\label{lapse} 
N=\frac{12\pi R_0^3e^{-3\Omega}}{H}
\end{equation} 
The lapse function will have the same form in the presence of a perfect fluid or of curvature. In order to find the isotropic equilibrium points of this equations system, it is necessary to partially rewrite it using some bounded variables. The form of the ADM Hamiltonian suggests us
defining: 
%-equation-%
\begin{equation} x=H^{-1} 
\end{equation} 
%-equation-% 
\begin{equation}
y=\pi R_0^3\sqrt{e^{-6\Omega}U}H^{-1} 
\end{equation} 
%-equation-%
\begin{equation} 
z=p_{\phi}\phi(3+2\omega)^{-1/2}H^{-1} 
\end{equation}
These new variables have all a physical interpretation:
\begin{itemize} 
\item the variable $x^2$ is proportional to the shear parameter $\Sigma$ introduced into \cite{WaiEll97}.
\item the variable $y^2$ is proportional to $(\rho_\phi-p_\phi)/(d\Omega/dt)^2$, $(d\Omega/dt)^2$ representing the Hubble function when the Universe is isotropic. 
\item the variable $z^2$ is proportional to $(\rho_\phi+p_\phi)/(d\Omega/dt)^2$. 
\item Defining the scalar field density parameter $\Omega_\phi$ like the ratio of its energy density with the square of the Hubble function, we deduce from the last two points that 
\begin{itemize} 
\item $\Omega_\phi$ is a linear combination of $y^2$ and $z^2$.
\item $\Omega_\phi$ is proportional to $y^2$ and $z^2$ when the scalar field is quintessent. 
\end{itemize} 
\end{itemize} 
One can check that these variables are bounded by rewriting the Hamiltonian constraint in the form of a sum of their squares: 
%-equation-%
\begin{equation}\label{BISMcont} 
p^2x^2+24y^2+12z^2=1 
\end{equation}
with $p^2=p_+^2+p_-^2$. So that they be real, we will consider that $3+2\omega$ and $U$ are positive functions. The first condition is necessary to the respect of the weak energy condition which states that $\rho_\phi+p_\phi>0$. The Hamiltonian equations can be rewritten according to these variables in the following first order differential equations system: 
%-equation-%
\begin{equation}\label{BISM1} 
\dot{x}=72y^2x 
\end{equation}
%-equation-% 
\begin{equation} 
\dot{y}=y(6\ell z+72y^2-3)
\end{equation} 
%-equation-% 
\begin{equation}\label{BISM3}
\dot{z}=24y^2(3z-\frac{1}{2}\ell) 
\end{equation} 
%-equation-%
\begin{equation}\label{BISM4}
\dot{\phi}=12z\frac{\phi}{\sqrt{3+2\omega}} 
\end{equation} 
where $\ell$ is a function of the scalar field defined by:
$$
\ell=\phi U_\phi U^{-1} (3+2\omega)^{-1/2} 
$$ 
We thus reduced the seven Hamiltonian equations in the form of a system of four equations with
four variables $x$, $y$, $z$ and $\phi$, this last one not being necessarily bounded. This reduction of the equations number comes owing to the fact that the equations of Hamilton show that $p_{\pm}$ are some constants and that $\beta_+\propto \beta_-$. One thus eliminates three from the seven equations of Hamilton. Our next objective is then to find the equilibrium points corresponding to a stable isotropic state for the Universe and whose properties are defined in the following section
%--------------------------------------------------------------------------------------------------------------------------------------------------------------------%
\subsubsection{Definition of a stable isotropic state}\label{s2122}
Isotropy is defined in the article of Collins and Hawking\cite{ColHaw73} when the proper time diverges using the four following properties: 
\begin{itemize} 
\item
$\Omega\rightarrow  -\infty$\\
This condition says to us that the Universe is forever expanding when isotropisation occurs. Considering that no period of contraction was observed since radiation-matter decoupling  and that our Universe is strongly isotropic, this appears justified.
\item Let $T_{\alpha\beta}$ be the energy-momemtum tensor: $T^{00}>0$ and $\frac{T^{0i}}{T^{00}}\rightarrow  0$\\
$\frac{T^{0i}}{T^{00}}$ represents a mean velocity of the matter compared to surfaces of homogeneity. If this quantity did not tend to zero, the Universe would not appear homogeneous and isotropic.
\item Let be $\sigma_{ij}=(de^\beta/dt)_{k(i}(e^{-\beta})_{j)k}$ and $\sigma^2=\sigma_{ij}\sigma^{ij}$:
$\frac{\sigma}{d\Omega/dt}\rightarrow  0$.\\
This condition says that the anisotropy measured locally through the constant of Hubble $H_0$ tends to zero. Indeed, when we measure the Hubble constant, we evaluate the quantity $H_i=\frac{dg_{ii}}{dt}/g_{ii}=d\beta_{ii}/dt-d\Omega/dt$. So that it appears to be the same one in all the directions, it is thus necessary that $d\beta_{ii}/dt<<d\Omega/dt$.
\item $\beta$ tends to a constant $\beta_0$\\
This condition is justified by the fact that the anisotropy measured in the CMB is to some extent a measurement of the change of the matrix $\beta$ between time when radiation was emitted and time when it was observed. If $\beta$ did not tend to a constant, one would expect large quantities of
anisotropies in some directions.\\
\end{itemize} 
Within the framework of the Bianchi type $I$ model, we have that $d\beta_\pm/dt=-N^{-1}\dot\beta_\pm\propto e^{3\Omega}$. Hence, so that $\beta_\pm$ tend asymptotically to a constant in accordance with the fourth point, it is necessary that $\Omega\rightarrow -\infty$ in agreement with the first one and then, $d\beta_\pm/dt$ vanish. However, generally, when the derivative of a function tends to zero in $t\rightarrow  +\infty$, this necessarily does not imply that the function tends to a constant. It is for example the case of the logarithm $\ln t$. Hence the last point in Collins and Hawking definition implies that the isotropy appears relatively quickly. Note that the vanishing of $d\beta_\pm/dt$ is necessary so that the Universe be isotropic otherwise, $H_1-H_2$ as instance will be different from zero which is not observed. This is an important point because in presence of curvature, we will see that the limit $\Omega\rightarrow -\infty$ is not sufficient so that $d\beta_\pm/dt$ vanish whereas here it is essential and is a better justification than the convergence of $\beta_\pm$ to a constant. Concerning the third point, it means that the shear parameter proportional to $x\propto \dot\beta_\pm =\frac{d\beta_\pm}{dt}\frac{dt}{d\Omega}$, tends to zero. Consequently, the stable isotropic equilibrium points that we are looking for will be such as: 
$$
\Omega\rightarrow  -\infty 
$$
$$ 
x\rightarrow  0 
$$ 
By examining the equations system (\ref{BISM1}-\ref{BISM3}), one notes that there are three types of isotropic equilibrium states corresponding to these two limits and making it possible to define three isotropisation classes in the following way: 
\begin{itemize} 
\item Class 1: all the variables $(x,y, z)$ reach an isotropic equilibrium state with $y\not=0$. It is the class which seems to correspond to the most studied scalar-tensor theories in the literature.
\item Class 2: all the variables $(x,y, z)$ reach an isotropic equilibrium state with $y=0$. In this case, it is generally not possible to determine the asymptotic behavior of $x$ with the approach of the isotropy because it depends on the unknown way in which $y$ tends to zero whereas it is $x$ which determine the common asymptotic behavior of the metric functions at the isotropisation time. The class 2 isotropisation was observed numerically for a non minimally scalar field in \cite{a4}. 
\item Class 3: $x$ tends to equilibrium but not necessarily the other variables. As they must be bounded when $\Omega\rightarrow  -\infty$, that means that they have to oscillate in such a way that their first derivative with respect to $\Omega$ oscillate around zero. The class 3 isotropisation was observed numerically for complex scalar fields in \cite{a3}. 
\end{itemize} 
In this work our attention will be concerned with the study of class 1 isotropisation.
%--------------------------------------------------------------------------------------------------------------------------------------------------------------------%
\subsubsection{Study of the isotropic equilibrium states}\label{s2123} 
This section breaks up into three parts. In the first one, we calculate the location of the isotropic equilibrium points and in the second one we describe the assumptions which will enable us to derive the asymptotic behaviors of the metric functions and potential in the vicinity of the isotropy in the third section.\\\\ 
\underline{equilibrium points}\\\\ 
The equilibrium points of the system (\ref{BISM1}-\ref{BISM3}) such as $x=0$ and $y\not=0$ are given by: 
%-equation-%
\begin{equation}\label{BISMpeq} 
(x, y, z)=(0,\pm(3-\ell^2)^{1/2}/(6\sqrt{2}), \ell/6) 
\end{equation} 
In order to reach equilibrium, it is thus necessary that $\ell^2$ tends to a constant smaller than three. Linearising the equation (\ref{BISM1}) in the vicinity of equilibrium, it is found that asymptotically when $\Omega\rightarrow  -\infty$, the variable $x$ behaves like: 
%-equation-%
\begin{equation}\label{xeq} 
x\rightarrow  x_0e^{3\Omega-\int\ell^2d\Omega} 
\end{equation} 
\underline{Assumptions and stability of the asymptotic behaviors}\\\\ 
Before further going, this first calculation (\ref{xeq}) of an asymptotic behavior offers the opportunity to speak about the stability of our results. They can be separate in three types: 
\begin{enumerate}
\item The location of the isotropic equilibrium points. 
\item The conditions necessary to their existence. 
\item The asymptotic solutions associated to the equilibrium points. 
\end{enumerate} 
However the \emph{asymptotical} solutions and some of the necessary conditions will be determined by calculating the \emph{exact} solutions for each equilibrium point and they will be correct only if on one hand $\ell$ and in the other hand the variables $(y,z)$ tend sufficiently quickly to their equilibrium values. Otherwise, they will be different and we are going to explain why.
\\\\
With regard to $\ell$, in order to obtain some of our results in closed form and comparable between them, we will generally make the following assumption that we will call "variability assumption of $\ell$": 
\begin{itemize} 
\item When near equilibrium $\ell^2$ tends to a constant $\ell_0^2$, vanishing or not, with a variation $\delta\ell^2$ such as $\ell^2\rightarrow  \ell_0^2+\delta\ell^2$, $\int\ell^2 d\Omega\rightarrow \ell_0^2\Omega+const$ (If $\ell_0=0$, the integral thus tends to a constant).
\end{itemize} 
We will see in the third part of this section that it is possible to asymptotically determine the behavior of $\phi(\Omega)$ and thus of $\ell(\Omega)$. Consequently, it is possible to avoid any approximation with respect to $\ell$ as calculation (\ref{xeq}) shows it above: whatever the speed to which $\ell$ tends to its equilibrium value, the presence of the $\int\ell^2d\Omega$ term makes it possible to take into account the variation $\delta\ell^2$ of $\ell^2$ in its neighborhood. In order to show that one can mathematically free oneself from this assumption, all the results of the section \ref{s2} will be expressed by taking into account the integral of $\ell^2$ and then the variability assumption of $\ell$. Moreover, we will apply these results to the cases of two scalar-tensor theories, respectively in agreement and disagreement with this assumption.\\\\ 
With regard to $y$ and $z$, we  will make the same type of assumptions as for $\ell$ by considering that their variations $\delta y$ and $\delta z$ when we approach the equilibrium are sufficiently small to be negligible. For example, when we calculate the asymptotical behavior (\ref{xeq}) for $x$, we take into account the way in which $\ell$ approach its asymptotic value since we do not make the variability assumption of $\ell$ but not that of $y$ which we simply replaced by its equilibrium value in the field equations without considering $\delta y$. This problem can not be solved as "easily" as the one in connection with $\ell$. It is possible that a perturbative study can bring some answers but it is not guaranteed because it could strongly depend on the specification of the forms of $\omega$ and $U$ as functions of the scalar field.
\\\\
To summarize, all our results implying the calculation of an asymptotic approach of a quantity in the vicinity of equilibrium will be valid when the Universe reaches sufficiently quickly the isotropic state. The assumption on $\ell$ can be raised while not making the variability assumption but that seems more difficult for the variables $y$ and $z$. Hence, we will systematically assume that these variables sufficiently quickly approach their equilibrium values. These assumptions will be also valid for the variables $k$ and $w$ that we will define later, respectively associated with the presence of a perfect fluid and curvature.
\\\\ 
\underline{Asymptotical behaviors}
\\\\ 
Applying the variability assumption of $\ell$ to (\ref{xeq}) we have that when $\ell$ tends to a non vanishing constant, $x\rightarrow  e^{(3-\ell^2)\Omega}$ and to $e^{3\Omega}$ otherwise. This variable thus disappears when $\Omega\rightarrow -\infty$ and the reality condition of the equilibrium points is respected.\\
At the same time, the equation (\ref{BISM1}) for $\dot x$ shows that $x$ is a monotonous function of $\Omega$: when $x$ is initially positive (negative), it is asymptotically increasing (decreasing). $x$ is thus also of constant sign. Using the expression (\ref{lapse}) for the lapse function $N$ and owing to the fact that $dt=-Nd\Omega$, we deduce that $\Omega(t)$ is a decreasing (increasing) function of the proper time $t$ when $x$ , or in an equivalent way the Hamiltonian, is initially positive (negative). Consequently, an initially positive Hamiltonian is a necessary initial condition so that isotropy when $\Omega\rightarrow  -\infty$ occurs at late times. Finally, a last remark about the monotonous functions. We can calculate that $dg_{ij}/d\Omega=-2e^{-2\Omega+\beta_{ij}}(1-\dot\beta_{ij})$. Taking into account what we said on the monotony of $x$ and the expression for the derivatives of $\beta_{ij}$ with respect to $\Omega$, it comes that the $\beta_{ij}$ are monotonous functions of proper time $t$ and that consequently each metric function can not have more than one extremum during time evolution. We had shown this point in a different way in \cite{Fay00A} by using the ADM Hamiltonian formalism.\\
To determine $\phi(\Omega)$, we use the equation (\ref{BISM4}) for $\dot\phi$ which is written asymptotically: 
$$ 
\dot\phi=2\frac{\phi^2U_\phi}{U(3+2\omega)} 
$$ 
In this last expression the variability assumption of $\ell$ is not made but on the other hand we neglect the variation $\delta z$ of $z$ in the isotropy neighborhood. It is the asymptotic form of the solution of this equation which will give us the asymptotic behavior of $\phi$ according to $\Omega$. One will thus deduce $\ell(\Omega)$ and $U(\Omega)$ which are two unknown functions of the scalar field. In particular, to know $\ell(\Omega)$ will make it possible to check the necessary conditions for isotropy when $\Omega\rightarrow  -\infty$, the variability assumption of $\ell$ and to calculate the asymptotic form of the metric functions $\Omega(t)$ and the potential $U(t)$.  Indeed, using on the one hand the asymptotic behavior of $x$ and the relation $dt=-Nd\Omega$ and on the other hand the definition of $y$ , it is found that with the approach of the isotropy $\Omega$ and $U$ behave respectively like: 
$$
dt=-12\pi R_0^3x_0e^{-\int \ell^2d\Omega}d\Omega 
$$
and
$$ 
U=e^{2\int\ell^2 d\Omega} 
$$ 
and then, if we take into account of the variability assumption of $\ell$ 
$$ 
dt=-12\pi R_0^3x_0e^{-\ell^2\Omega}d\Omega 
$$ 
and 
$$ 
U=e^{2\ell^2 \Omega}
$$ 
This assumption enables us to calculate that when $\ell^2$ tends to a non vanishing constant, the metric functions tend to $t^{\ell^{-2}}$ and the potential to $t^{-2}$. On the other hand, when $\ell^2$ tends to vanish, the Universe tends to a De Sitter model and the potential to a cosmological constant. If the assumption is not checked, it is always possible to determine these asymptotic behaviors but with some quadratures.
%--------------------------------------------------------------------------------------------------------------------------------------------------------------------%
\subsubsection{Discussion and applications}\label{s2124} 
\begin{figure}[h ] 
\centering
\includegraphics[width=\textwidth]{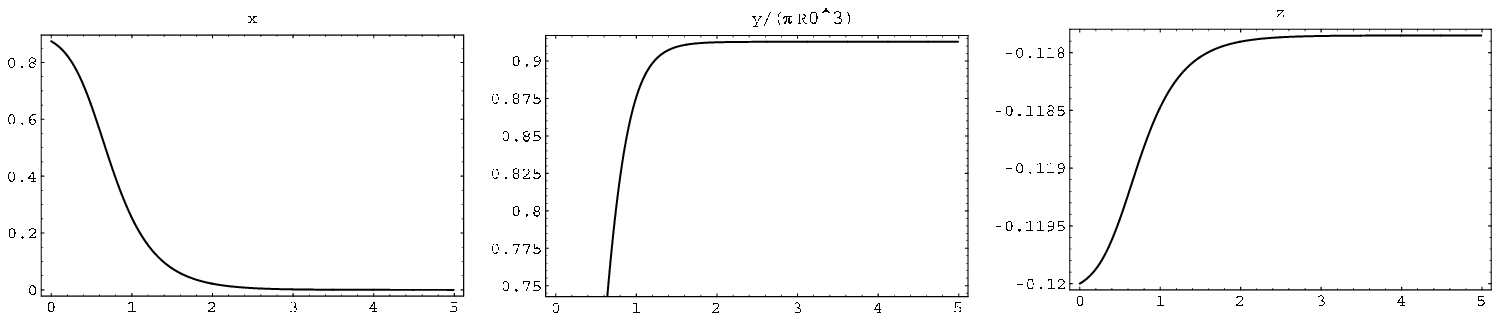}
\caption{\scriptsize{\label{BISMfig1}Evolution of the variables $x$, $y$ and $z$ when $\frac{(3+2\omega)^{1/2}}{\phi}=\sqrt{2}$, $U=e^{m\phi}$, $R_0^3=(\sqrt{24}\pi)^{-1}$ and $m=-1$ with the initial values $(x, y, z, \phi)=(0.87, 0.25, -0.12, 0.14)$. $\ell=-1/\sqrt{2}$, $x$ tends to $0$, $y(\pi R_0^3)^{-1}$ to $\sqrt{3-\ell^2}/(6\pi R_0^3\sqrt{2})=0.91$ and $z$ to $\ell/6=0.12$ in agreement with the equilibrium points expression.}} 
\end{figure}

\begin{figure}[h ] 
\centering
\includegraphics[width=\textwidth]{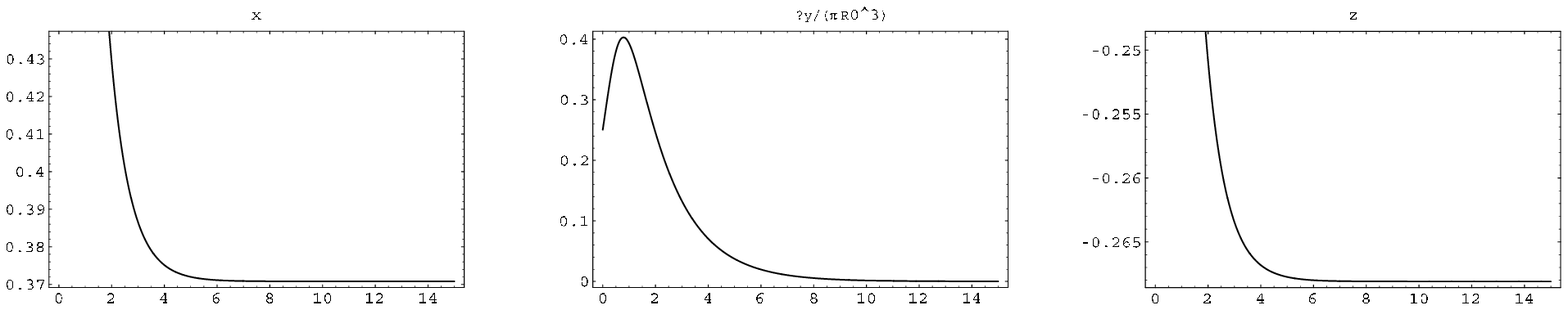}
\caption{\scriptsize{\label{BISMfig3}Evolution of the variables $x$, $y$ and $z$ when $\frac{(3+2\omega)^{1/2}}{\phi}=\sqrt{2}$, $U=e^{m\phi}$, $R_0^3=(\sqrt{24}\pi)^{-1}$ and $m=-3.2$ with the initial values $(x, y, z, \phi)=(0.87, 0.25, -0.12, 0.14)$. Universe does not isotropise:
the system tends to an anisotropic equilibrium point such as the functions $\beta_\pm$ describing the anisotropy diverge.}}
\end{figure}

\begin{figure}[h ]
\centering \includegraphics[width=\textwidth]{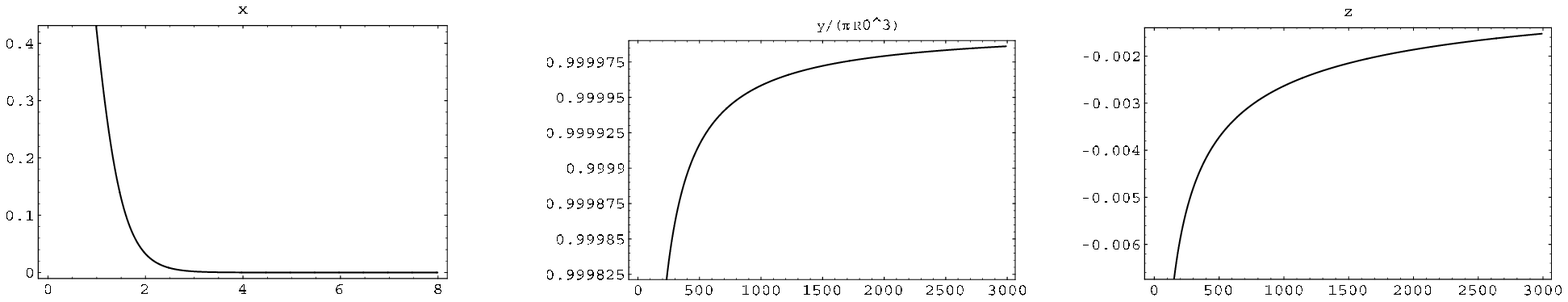}
\caption{\scriptsize{\label{BISMfig2}Evolution of the variables $x$, $y$ and $z$ when $\frac{(3+2\omega)^{1/2}}{\phi}=\sqrt{2}$, $U=\phi^m$, $R_0^3=(\sqrt{24}\pi)^{-1}$ and $m=-1$ with the initial values $(x, y, z, \phi)=(0.87, 0.25, -0.12, 0.14)$. $x$ tends to $0$, $y(\pi R_0^3)^{-1}$ to $1$ and $z$ to $0$ in agreement with the equilibrium points expression. Let us notice that the variables $y$ and $z$ tend much less quickly to their equilibrium values than on the figure \ref{BISMfig3}. This is due to the "slowness" of the convergence of $\ell$ to zero which then reflects on the variation of these variables.}} 
\end{figure}

The results of section \ref{s212} are concerned by a minimally coupled and massive scalar-tensor theory without a perfect fluid. It is the simplest of the theories that we will consider and the method used above will serve us as a guide for the next sections. We have assumed that $U$ and $3+2\omega$ are positive functions and a class 1 isotropisation. When it is supposed that the function $\ell$ and the variables $y$ and $z$ tend sufficiently quickly to their equilibrium values, we have the following results:\\\\ 
\emph{
Let us consider a minimally coupled and massive scalar-tensor theory and the function $\ell$ defined by $\ell=\frac{\phi U_\phi}{U(3+2\omega)^{1/2}}$. The asymptotic behavior of the scalar field to the approach of the isotropy is given by the form of the solution when $\Omega\rightarrow  -\infty$ of the differential equation $\dot\phi=2\frac{\phi^2 U_\phi}{U(3+2\omega)}$. A necessary condition for class 1 isotropisation is that $\ell^2<3$. If $\ell$ tends to a non vanishing constant, the metric functions tend to $t^{\ell^{-2}}$ and the potential disappears like $t^{-2}$. If $\ell$ tends to vanish, the Universe tends to a De Sitter model and the potential to a constant.}
\\\\ 
The variability assumption of $\ell$ can be raised and the results are expressed then using the integral of $\ell^2$ as shown above. They are in agreement with the "Cosmic No Hair theorem" of Wald\cite{Wal83} which says that homogeneous and initially expanding models with a positive cosmological constant (except the Bianchi type $IX$ model) and an energy-momemtum tensor satisfying the strong and dominant energy conditions, tend to an isotropic De-Siter model for which the expansion is exponential. Here, when a cosmological constant is considered or when $\ell\rightarrow  0$ such as the variability  assumption of $\ell$ is checked, the Universe, when it isotropises, tends to a De Sitter model and the potential to a constant. On the other hand, when $\ell$ tends to vanish but the variability assumption of $\ell$ is not checked, the potential does not tend any more to a constant and the Universe does not approach a De Sitter model.\\\\ 
As an application, we will examine the cases of the scalar-tensor theories defined by the Brans-Dicke coupling function
$$
\frac{(3+2\omega)^{1/2}}{\phi}=\sqrt{2}
$$ and the forms of
potentials 
$$ 
U=e^{m\phi} 
$$ 
and 
$$ 
U=\phi^m 
$$ 
It is pointed out that some of our results represent necessary conditions and that consequently,
when in the applications below we will speak about isotropisation, it is always by keeping in mind that they could not be sufficient.\\ 
The exponential potential of $\phi$ was largely used in the literature. Isotropisation of Bianchi models with this potential was already studied in \cite{ColIbaHoo97} and thus will enable us to test our results. It was shown that all the Bianchi models (except the Bianchi type $IX$ model when it contracts) isotropised at late times when $m^2<2$. If $m=0$ , the Universe tends to a De Sitter model because the potential is a constant and, if not, it is expanding such as $e^{-\Omega}\rightarrow t^{2m^{-2}}$. If $m^2>2$ , the Bianchi type $I$, $V$, $VII$ and $IX$ models may isotropise at late times. Now, using our results, we get that asymptotically: 
$$  
\phi\rightarrow  m\Omega 
$$ 
The necessary condition for class 1 isotropisation is then $m^2<6$ and the asymptotic behaviors of the metric functions are well in agreement with what was predicted in \cite{ColIbaHoo97}. The difference between the results of this last paper and ours is the nature of the range of $m$ authorizing the isotropisation since we find an upper limit. The figure \ref{BISMfig1} illustrates the convergence of the variables $x$, $y$ and $z$ to their equilibrium values for $m=-1$.
When $m^2>6$ , the class 1 isotropisation is not possible any more because $y$ equilibrium value would be complex. A numeric simulation of this case is shown on the figure \ref{BISMfig3}: when $m=-3.2$, the Universe always tends to an equilibrium state which is now anisotropic because $x$ tends to a non vanishing constant and thus the functions $\beta_\pm$ describing the isotropy diverge. 
\\
Now let us examine the case of a power law potential of the scalar field. Then we get for $\ell$:
$$ 
\ell\rightarrow \frac{m}{\sqrt{2}\phi} 
$$ 
and, if a class 1 isotropisation occurs, the scalar field behaves as 
$$
\phi^2\rightarrow  2m\Omega
$$ 
We must thus have $m<0$ so that the scalar field is real and we deduce that $\ell^2$ tends to
vanish like $m(4\Omega)^{-1}$. Consequently, $\int\ell^2d\Omega$ does not tend to a constant when $\Omega\rightarrow -\infty$ but diverges like $\frac{m}{4}\ln (-\Omega)$ and we must take account of this integral in our results: the variability assumption of $\ell$ is not checked here. Raising this assumption, one finds then that the potential tends to vanish like $(-\Omega)^{m/2}$ and the metric functions tend to $exp\left[(\frac{4-m}{48\pi>R_0^3 x_0}t)^{\frac{4}{4-m}}\right]$. So that this quantity diverges positively, it is thus necessary that $m<4$ what is always checked since $m<0$. This case is illustrated on the figure \ref{BISMfig2} where it is seen very clearly that convergence of the variables $y$ and $z$ to their equilibrium values is slow(compare with the figure \ref{BISMfig1} of the previous application). This means that the Universe approaches "slowly" its isotropic state and one could then think that, in addition to raising the variability assumption of $\ell$, the variations $\delta y$ and $\delta z$ of the variables $y$ and $z$ near equilibrium about which we speak in the previous sub-section should also be taken into account. However, it seems that these last corrections are not necessary. This can for instance be checked by comparing the asymptotic evolution of $z$ corresponding theoretically to $z\rightarrow \ell/6$  with the numerical integration of the figure \ref{BISMfig2} for the large $\Omega$ values. 
\\ 
When $m>0$ , a class 1 isotropisation does not seem possible any more because then the scalar field would be complex. Numerical integrations show that the initially positive scalar field decreases to zero for a finite $\Omega$ value. Then they fail for negative value, maybe indicating a singularity occurrence.
%--------------------------------------------------------------------------------------------------------------------------------------------------------------------%
\subsection{Perfect fluid case}\label{s213}
The method is the same one as without a perfect fluid but an additional term appears in the field equations\cite{Fay01A} due to its presence. Its equation of state is $p=(\gamma-1)\rho$ with $\gamma\in\left[1,2\right]$. We will also consider that the pressure of the non tilted perfect fluid is isotropic. It is a simplifying assumption whose consequence for the Bianchi type $I$ model could be to increase the anisotropies decay(in $V^{-1}$), thus preventing any observational detection, whereas the presence of an anisotropic pressure slow down it. In this last case, the anisotropy could then be detected via the ratio $\delta T/T$ of the CMB which depends on the shear parameter at the time of last scattering surface\cite{Bar97A}.
%--------------------------------------------------------------------------------------------------------------------------------------------------------------------%
\subsubsection{Field equations}\label{s2131} 
This time the ADM Hamiltonian is written: 
%-equation-%
\begin{equation} \label{hamiltonian} 
H^2 = p_+ ^2 +p _ - ^2+12\frac{p_\phi ^2 \phi^2}{3+2\omega}+24\pi^2 R_0 ^6e^{-6\Omega}U+\delta e^{3(\gamma-2)\Omega} 
\end{equation} 
where  $\delta$ is a positive constant. Compared to the previous section, one thus sees appearing the term $\delta e^{3(\gamma-2)\Omega}$ due to the presence of the perfect fluid. The Hamiltonian equations become: 
$$  
\dot{\beta}_\pm = \frac{\partial H}{\partial p_\pm}=\frac{p_\pm}{H} 
$$ 
$$ 
\dot{\phi}=\frac{\partial H}{\partial p_\phi}=\frac{12\phi^2 p_\phi}{(3+2\omega)H} 
$$
$$ 
\dot{p}_\pm=-\frac{\partial H}{\partial \beta_\pm}=0 
$$
$$ 
\dot{p}_\phi=-\frac{\partial H}{\partial \phi}=-12\frac{\phi p_\phi^2}{(3+2\omega)H}+12\frac{\omega_\phi \phi^2 p_\phi ^2}{(3+2\omega)^2H}-12\pi^2 R_0 ^6 \frac{e^{-6\Omega}U_\phi}{H} 
$$
$$ 
\dot{H}=\frac{dH}{d\Omega}=\frac{\partial H}{\partial \Omega}=-72\pi^2 R_0 ^6 \frac{e^{-6\Omega}U}{H}+3/2\delta(\gamma-2)\frac{e^{3(\gamma-2)\Omega}}{H}
$$ 
In order to rewrite these equations, we will use the same bounded variables $x$, $y$ and $z$ that previously defined to which we will add a fourth variable: 
$$  
k^2=\delta e^{3(\gamma-2)\Omega}H^{-2} 
$$ 
related to the presence of the perfect fluid. This variable is in fact proportional to the density parameter of the perfect fluid, one of the main cosmological parameters usually noted $\Omega_m$. This can be shown by checking that $k^2\propto V^{-\gamma}/(\frac{d\Omega}{dt})^2$ where $\frac{d\Omega}{dt}$ is the Hubble function when the Universe isotropises. $k$ is not independent of the other variables and can be rewritten like: 
%-equation-%
\begin{equation} 
k^2=\delta x^\gamma y^{2-\gamma}U^{\gamma/2-1}\nonumber 
\end{equation} 
%-equation-%
\begin{equation}\label{form1} 
k^2=\delta x^2e^{3(\gamma-2)\Omega}
\end{equation} 
%-equation-% 
\begin{equation}\label{form2} 
k^2=\delta y^2 U^{-1} V^{-\gamma} 
\end{equation} 
Then, the Hamiltonian constraint becomes:
%-equation-%
\begin{equation} \label{BIAMc} 
p^2x^2+R^2y^2+12z^2+k^2=1
\end{equation} 
and the Hamiltonian equations are: 
%-equation-%
\begin{equation} \label{BIAM1} 
\dot{x}=72y^2x-3/2(\gamma-2)k^2x
\end{equation} 
%-equation-% 
\begin{equation} \label{} 
\dot{y}=y(6\ell z+72y^2-3)-3/2(\gamma-2)k^2y \end{equation} 
%-equation-%
\begin{equation} \label{}
\dot{z}=24y^2(3z-\frac{\ell}{2})-3/2(\gamma-2)k^2z 
\end{equation}
%-equation-% 
\begin{equation}\label{BIAM4}
\dot{\phi}=12z\frac{\phi}{(3+2\omega)^{1/2}} 
\end{equation} 
with always $\ell=\phi U_\phi U^{-1}(3+2\omega)^{-1/2}$.
%--------------------------------------------------------------------------------------------------------------------------------------------------------------------%
\subsubsection{Study of the isotropic states}\label{s2132} 
One distinguishes two types of equilibrium states according to whether $k$ , i.e. the density parameter of the perfect fluid, tends to vanish or to a non vanishing constant.
\\\\ 
\underline{$k$ tends to vanish}
\\\\ 
As $y$ does not tend to vanish because we consider a class 1 isotropisation, we deduce from the form (\ref{form2}) of $k$ that $U>> V^{-\gamma}$. The equilibrium points are the same ones as in the
absence of perfect fluid and thus we find again the reality condition $\ell^2<3$. The asymptotic behavior of $x$ near equilibrium is obtained starting from the equation (\ref{BIAM1}):
%-equation-%
\begin{equation}\label{xeq1} 
x\rightarrow  x_0 e^{3\Omega-\int\ell^2 d\Omega-\frac{3}{2}(\gamma-2)\int k^2 d\Omega} 
\end{equation} 
From this expression and the definition of $y$, we then deduce for the potential: 
$$  
U\rightarrow  U_0 e^{2\int\ell^2d\Omega+3(\gamma-2)\int k^2 d\Omega} 
$$ 
While making use of the asymptotic behavior of $x$ and the definition (\ref{form1}) of $k$, it comes: 
$$  
k^2=\delta x_0^2 e^{-2\int\ell^2 d\Omega-3(\gamma-2)\int k^2d\Omega+3\gamma\Omega} 
$$ 
By deriving this expression, one obtains the differential equation 
$$ 
2k\dot k=\left[-2\ell^2-3(\gamma-2)k^2+3\gamma\right]k^2 
$$ 
whose exact solution is: 
$$ 
k^2=\frac{e^{3\gamma\Omega-2\int\ell^2d\Omega}}{k_0+3(\gamma-2)\int e^{3\gamma\Omega-2\int\ell^2d\Omega}d\Omega} 
$$ 
$k_0$ being an integration constant. All these results were obtained without applying the variability assumption of $\ell$. So now we take it into account and we obtain: 
$$ 
k^2\rightarrow \delta x_0^2e^{(-2\ell^2+3\gamma)\Omega} 
$$ 
It thus follows that $k\rightarrow  0$ when $\Omega\rightarrow  -\infty$ if $\ell^2<\frac{3\gamma}{2}<3$. Consequently the isotropisation will occur when $k\rightarrow 0$ only if $\ell^2<\frac{3\gamma}{2}$ which is a smaller range than when no perfect fluid is present(then $\ell^2<3$). However, the asymptotic behaviors of the metric functions and the potential remain unchanged but if $k$ sufficiently quickly does not approach its equilibrium value.\\ 
If the variability assumption is not valid, again one must take into account the integral of $\ell^2$. The range of $\ell^2$ allowing the isotropisation will be always at least such as $\ell^2<3$ because this inequality is independent of any approximation but it will be modified (or not) differently by the fact that we consider the limit $k\rightarrow 0$. Moreover the asymptotic behaviors of the metric functions and potential will be different from what they are when there is no perfect fluid in spite of this disappearance of $k$.
\\\\
\underline{$k$ tends to a non vanishing constant}
\\\\ 
The isotropic equilibrium points are not more the same ones and thus either the asymptotic behaviors of the metric functions and the potential. For the first ones, we find: 
$$  
(x, y, z)=(0,\pm\frac{\sqrt{\gamma(2-\gamma)}}{4\sqrt{2}\ell},\frac{\gamma}{4\ell}) 
$$ 
after having deduced from the constraint that 
$$ 
k^2=1-3\gamma(2\ell^2)^{-1} 
$$ 
The equilibrium points will be real if $\gamma$ is a positive constant smaller than $2$, in agreement with the range of variation of $\gamma$ which we specified, that is $\gamma\in\left[1,2\right]$. The variable $k$ will be real and the other variables will reach equilibrium for a not vanishing value of $y$, thus respecting the definition of class 1 isotropy, if $\ell^2$ tends to a constant larger than $\frac{3\gamma}{2}$. This necessary condition for isotropy is independent of any approximation. Linearising the differential equation for $x$ , we find that asymptotically: 
$$  
x\rightarrow  e^{\frac{3}{2}(2-\gamma)\Omega}
$$ 
and that the metric functions tend to
$$ 
e^{-\Omega}\rightarrow  t^{\frac{2}{3\gamma}} 
$$ 
From the definition of $y$, we deduce that the potential tends to vanish as $t^{-2}$ what is confirmed by the form (\ref{form2}) of $k$ which shows that asymptotically
$$  
U\propto V^{-\gamma} 
$$ 
in agreement with the asymptotic expressions of the potential and the metric functions with respect to the proper time $t$. This last expression makes it possible to determine the scalar field asymptotic form $\phi(\Omega)$ according to the one of the potential. Let us note that all these asymptotic behaviors are independent of the variability assumption of $\ell$.
%--------------------------------------------------------------------------------------------------------------------------------------------------------------------%
\subsubsection{Discussion and applications}\label{s2134} 
\begin{figure}[h ]
\includegraphics[width=\textwidth]{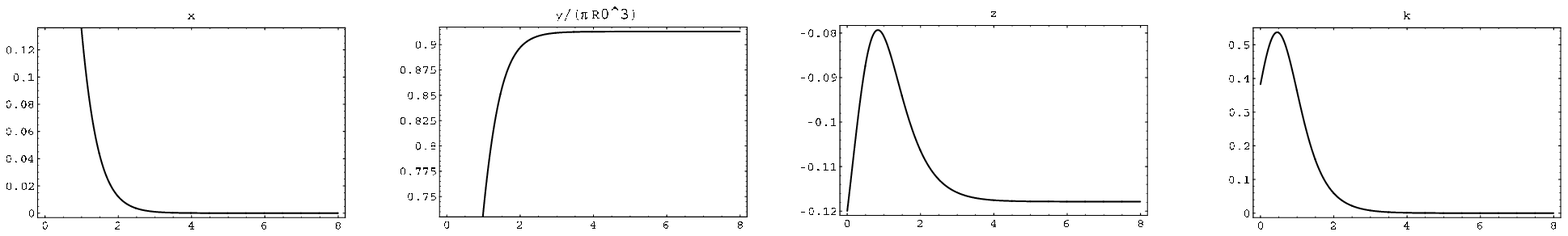}
\caption{\scriptsize{\label{BISMfig4}Evolution of the variables $x$ , $y$, $z$ and $k$ in the presence of a fluid of dust ($\gamma=1$) when $\frac{(3+2\omega)^{1/2}}{\phi}=\sqrt{2}$, $U=e^{m\phi}$, $R_0^3=(\sqrt{24}\pi)^{-1}$ and $m=1$ with the initial values $(x,y,z,\phi)=(0.62, 0.25, -0.12, -0.14)$. All occurs as in the vacuum case with $k$ (or equivalently $\Omega_m$) tending to zero.}} 
\end{figure}

\begin{figure}[h ] \includegraphics[width=\textwidth]{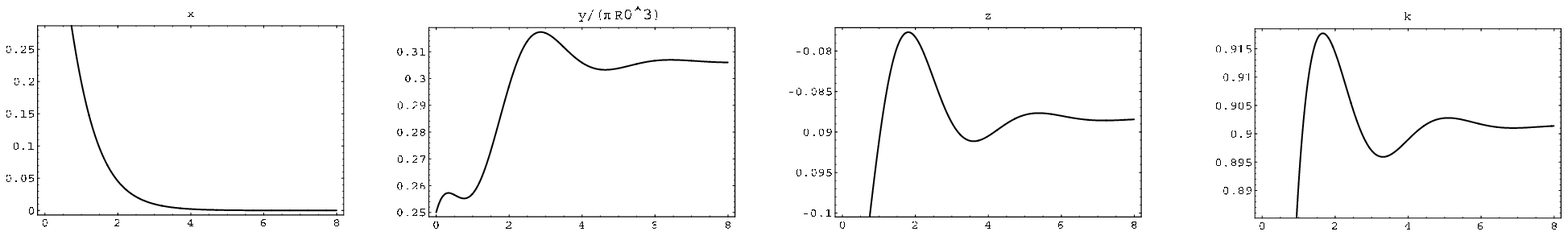}
\caption{\scriptsize{\label{BISMfig5}Evolution of the variables $x$, $y$, $z$ and $k$ in the presence of a fluid of dust ($\gamma=1$) when $\frac{(3+2\omega)^{1/2}}{\phi}=\sqrt{2}$, $U=e^{m\phi}$, $R_0^3=(\sqrt{24}\pi)^{-1}$ and $m=4$ with the initial values $(x,y,z,\phi)=(0.62, 0.25, -0.12, -0.14)$. $y$, $z$ and $k$ oscillate to their equilibrium values which are respectively $-0.0625$, $-0.088$ and $0.90$ when $\ell=-4/\sqrt{2}$.}}
\end{figure}

Let us summarize our results obtained in the presence of a perfect fluid. For that, we will
state them according to the density parameter $\Omega_m$ of the perfect fluid which is proportional to $k$. When it tends to vanish, we will make the variability assumption of $\ell$ otherwise it is useless as shown above. It comes:
\\\\ 
\emph{\underline{Isotropisation with $\Omega_m\rightarrow 0$}:\\ 
The results are the same ones as without a perfect fluid but the range of $\ell^2$ allowing the isotropisation is reduced to $\ell^2<\frac{3\gamma}{2}$. Moreover, at the time of the
isotropisation the scalar field potential is asymptotically larger than the perfect fluid energy density.}
\\\\ 
When the variability assumption of $\ell$ is not made, the things are not so simple and the results depend completely on the way in which $\ell^2$ approaches equilibrium. One can however always calculate them by using the expressions given in the preceding sections and depending on $\int\ell^2d\Omega$.\\\\ 
When $k$ tends to a non vanishing constant, the equilibrium state is different and we find that:\\\\ 
\emph{\underline{Isotropisation with $\Omega_m\not\rightarrow  0$}:\\ 
Let us consider a minimally coupled and massive scalar-tensor theory and the quantity $\ell$ defined by $\ell=\frac{\phi U_\phi}{U(3+2\omega)^{1/2}}$. The scalar field  asymptotic behavior at the approach of isotropy can be deduced owing to the fact that $U\propto V^{-\gamma}$: the potential of the scalar field is proportional to the perfect fluid energy density. A necessary condition for a class 1 isotropisation will be that $\ell^2>\frac{3\gamma}{2}$, $\ell^2$ is bounded and $0<\gamma<2$. Then the potential disappears like $t^{-2}$ and the metric functions tend to $t^{\frac{2}{3\gamma}}$.}\\\\   
This last result shows that the scalar-tensor theory tends to General Relativity with a perfect fluid in agreement with \cite{SanKalWag98}. Then, the presence of the scalar field have no effect on the asymptotic evolution of the metric functions. Thus for a fluid of dust such as $\gamma=1$, the Universe tends to that of Einstein-De Sitter with $e^{-\Omega}\rightarrow t^{2/3}$ and for a radiative fluid such as $\gamma=4/3$, to a Tolman Universe with $e^{-\Omega}\rightarrow t^{1/2}$. Also let us notice that the ranges of $\ell^2$ allowing isotropisation when $k\rightarrow 0$ and
$k\not\rightarrow 0$ are complementary.\\\\ 
Now we examine our results in the light of the two scalar-tensor theories which we had considered in the absence of a perfect fluid in the section \ref{s2124}.\\ 
Let us start by considering an exponential potential. The only difference with the results obtained without a perfect fluid is the necessary condition allowing the isotropisation. When $m^2<3\gamma$ , isotropisation should occur with $\Omega_m\rightarrow  0$, and the metric functions tend to $t^{2m^{-2}}$. This case is illustrated on the figure \ref{BISMfig4} for $m=1$. When $m^2>3\gamma$ , isotropisation should occur with $\Omega_m\not\rightarrow  0$, and the metric functions tend to $t^{\frac{2}{3\gamma}}$. This case is illustrated on the figure \ref{BISMfig5} for $m=4$. These results are in agreements with those found in \cite{CopLidWan98} for the FLRW models. However, in this last paper, a stable solution of trackers type such as $\Omega_m\rightarrow 0$ had also been found when $m^2>6$. Here, we do not find it since it does not allow for isotropy.\\ 
With regard to the power law potential when $\Omega_m\rightarrow 0$, we know that the variability assumption of $\ell$ is not verified since $\ell$ tends to zero but that the integral of its square diverges like $\frac{m}{4}\ln (- \Omega)$ when $\Omega\rightarrow -\infty$. Moreover, we showed that $m$ must be negative so that the scalar field be real. Taking into account these elements, the calculation of $k^2$ and its integral gives us then: 
$$  
k^2\rightarrow  k_0^2 (- \Omega)^{-m/2}e^{3\gamma\Omega}
$$
and 
$$
\int k^2d\Omega\propto \Gamma(1-m/2,-3\gamma\Omega)\rightarrow  0 
$$ 
when $\Omega\rightarrow  -\infty$, $\Gamma$ being the Euler function. Consequently, these two quantities also tend to zero when $\Omega\rightarrow -\infty$ without one needing additional condition. Simulations with $m=-1$ and a dust fluid give identical results to the figures \ref{BISMfig2}.\\ 
When $\Omega_m$ tends to a non vanishing constant, the scalar field behaves like $\phi\rightarrow 
e^{-\frac{3\gamma}{m}\Omega}$ and thus $\ell$ disappears or diverge, preventing a class 1 isotropisation.
%--------------------------------------------------------------------------------------------------------------------------------------------------------------------%
\section{Bianchi class $A$ models with curvature}\label{s3} 
In this section, we will consider the presence of curvature by studying the Bianchi class A models of types $II$, $VI_0$, $VII_0$, $VIII$ and $IX$ without a perfect fluid. This last model in particular, contains the solutions of the FLRW models with positive curvature.
%--------------------------------------------------------------------------------------------------------------------------------------------------------------------%
\subsection{Field equations}\label{s220} 
The ADM Hamiltonian for the Bianchi class $A$ models with curvature is written:
%-equation-%
\begin{equation} \label{hamiltonian} 
H^2 = p_+ ^2 +p _ - ^2+12\frac{p_\phi ^2 \phi^2}{3+2\omega}+24\pi^2 R_0 ^6 e^{-6\Omega}U+V(\Omega, \beta_+, \beta _-) 
\end{equation}
where $V(\Omega, \beta_+, \beta_-)$ is the curvature potential characterizing each Bianchi model and is given in table \ref{tab1}. The Hamiltonian equations are then: 
%-equation-%
\begin{equation} \label{bianBeta} 
\dot{\beta} _ \pm = \frac{\partial H}{\partial p _ \pm}=\frac{p_\pm}{H} 
\end{equation} 
%-equation-%
\begin{equation} \label{bianPhi} 
\dot{\phi}=\frac{\partial H}{\partial p_\phi}=\frac{12\phi^2 p_\phi}{(3+2\omega)H} 
\end{equation}
%-equation-% 
\begin{equation} \label{bianP}
\dot{p}_\pm=-\frac{\partial H}{\partial \beta _ \pm}=-\frac{\partial V}{2H\partial \beta _ \pm} 
\end{equation} 
%-equation-%
\begin{equation} \label{bianPphi} 
\dot{p}_\phi=-\frac{\partial H}{\partial \phi}=-12\frac{\phi p_\phi^2}{(3+2\omega)H}+12\frac{\omega_\phi \phi^2 p_\phi ^2}{(3+2\omega)^2 H}-12\pi^2 R_0 ^6 \frac{e^{-6\Omega}U_\phi}{H} 
\end{equation}
%-equation-% 
\begin{equation} \label{bianH}
\dot{H}=\frac{dH}{d\Omega}=\frac{\partial H}{\partial \Omega}=-72\pi^2 R_0 ^6 \frac{e^{-6\Omega}U}{H}+\frac{\partial V}{2H\partial\Omega}
\end{equation} 
As for the Bianchi type $I$ model, the isotropy means that $d\beta_\pm/d\Omega$ have to vanish. But from now on the conjugate moments of the $\beta_\pm$ functions are not any more some constants and thus $d\beta_\pm/dt=p_\pm e^{3\Omega}$: contrary to the Bianchi type $I$ model, the limit $\Omega\rightarrow -\infty$ does not imply any more that $d\beta_\pm/dt\rightarrow 0$ and thus that isotropy occurs for a forever expanding Universe. We have to show it. Let us suppose that isotropisation leads to a static Universe, i.e. such as $\Omega\rightarrow  const$, when the proper time diverges. Then, so that $d\beta_\pm/dt$ disappear, it is necessary that $p_\pm\rightarrow 0$. However the equations (\ref{bianP}) indicates that $dp_\pm/dt\propto -\frac{\partial V}{\partial \beta_\pm}e^{3\Omega}$. Consequently if $\Omega$ and $\beta_\pm$ tend to some constants when $t\rightarrow +\infty$, for the Bianchi types $II$, $VI_0$ and $VIII$ models, these derivative tend to non vanishing constants and the conjugate momenta $p_\pm$ can not disappear and the isotropy to occur. On the other hand, the things are not so simple for Bianchi type $VII_0$ and $IX$ models because if $\beta_\pm\rightarrow 0$, it is the same for $\frac{\partial V}{\partial \beta _ \pm}$ and one can say nothing on the asymptotic values of $p_\pm$. We will further show for these models also, that the isotropy can emerge only for a diverging value of $\Omega$.\\ 
Consequently, for the Bianchi models with curvature, isotropisation can occur only if: 
$$
\Omega\rightarrow \pm\infty
$$
$$
\frac{d\beta_\pm}{d\Omega}\rightarrow 0
$$
$$
p_\pm e^{3\Omega}\rightarrow  0
$$ 
We will see that the third limits are always satisfied when it is also the case for the two first ones. In what follows, the variability assumption of $\ell^2$ will be systematically applied. We did not explore what occurs when that it is raised. Indeed, we will see that the results obtained for the isotropisation of models with curvature are similar to those obtained for a flat model. However the conditions to check to show that the isotropy is reached are much more numerous and the variability assumption can not be raised easily. Let us note however that this is technically feasible in the same way as for the Bianchi type $I$ model.\\ 
In order to describe the curvature of the Bianchi models, we will introduce new variables prefixed $w$ and similar to the three $N_i$ variables defined by arguments of symmetry of the structure constants in \cite{WaiEll97} and \cite{RosJan88}.
%--------------------------------------------------------------------------------------------------------------------------------------------------------------------%
\subsection{Bianchi type $II$ model}\label{s2211}
In order to rewrite the field equations, we use the following variables: 
%-equation-%
\begin{equation} \label{BIIv1} 
x_\pm=p_\pm H^{-1} 
\end{equation}
%-equation-% 
\begin{equation} \label{BIIv2} 
y=\pi R_0^3\sqrt{U}e^{-3\Omega}H^{-1} 
\end{equation} 
%-equation-%
\begin{equation} \label{BIIv3} 
z=p_{\phi}\phi(3+2\omega)^{-1/2}H^{-1}
\end{equation} 
%-equation-% 
\begin{equation} \label{BIIv4} 
w=\pi R_0^2e^{-2\Omega+2(\beta_++\sqrt{3}\beta_-)}H^{-1} 
\end{equation} 
Only  one $w$ variable is sufficient to describe the curvature just as only one variable $N_i$ was sufficient in \cite{WaiEll97}. Then the condition $\frac{d\beta_\pm}{d\Omega}\rightarrow  0$, necessary to the isotropisation, writes $x_\pm\rightarrow 0$. It will be the same one for all the Bianchi models for whom we will re-use the same variables $x_\pm$, $y$ and $z$. The Hamiltonian constraint and the field equations rewrite like: 
%-equation-%
\begin{equation} \label{BIIcontrainte} 
x_+^2+x_-^2+24y^2+12z^2+12w^2=1
\end{equation} 
%-equation-% 
\begin{equation} \label{BIIeq1}
\dot{x}_+=72y^2x_++24w^2x_+-24w^2 
\end{equation} 
%-equation-%
\begin{equation} \label{BIIeq2} 
\dot{x}_-=72y^2x_-+24w^2x _--24\sqrt{3}w^2 
\end{equation} 
%-equation-% 
\begin{equation}
\label{BIIeq3} \dot{y}=y(6\ell z+72y^2-3+24w^2) 
\end{equation}
%-equation-% 
\begin{equation} \label{BIIeq4}
\dot{z}=y^2(72z-12\ell)+24w^2z 
\end{equation} 
%-equation-%
\begin{equation} \label{BIIeq5}
\dot{w}=2w(x_++\sqrt{3}x_-+12w^2+36y^2-1) 
\end{equation} 
The constraint shows that the variables (\ref{BIIv1}-\ref{BIIv4}) are bounded. Moreover, we find the usual equation for the scalar field: 
$$ 
\dot \phi=\frac{12\phi}{\sqrt{3+2\omega}}z
$$ 
In order to determine the asymptotic behavior of the metric functions and potential, we will need to know the asymptotic behavior of the Hamiltonian. We thus rewrite the Hamiltonian equation for $H$ in the form: 
%-equation-%
\begin{equation}\label{BIIhamp1} 
\dot{H}=-H(72y^2+24w^2)
\end{equation} 
It shows that $H$ is a monotonous function keeping its initial sign during its evolution. Consequently, one deduces from the lapse function that when $H$ is initially positive (negative),
$\Omega\rightarrow -\infty$ corresponds to late (respectively early) times and vice versa when
$\Omega\rightarrow +\infty$.\\\\ 
With these equations, we can from now on calculate the equilibrium points of the equations
(\ref{BIIeq1}-\ref{BIIeq5}). It comes:\\
\begin{itemize} 
\item $(y,w)=(0, 0)$ 
\item $(x_+, x_-, y, z, w)=(1, \sqrt{3}, 0, 0, \pm i/2)$
\item $(x_+, x_-, y, z, w)=(0, 0,\frac{\pm\sqrt{3-\ell^2}}{(6\sqrt{2})}, \ell/6, 0)$ 
\item $(x_+, x_-,y, z, w)=(\frac{\ell^2-1}{\ell^2+8},\sqrt{3}\frac{\ell^2-1}{\ell^2+8}, \pm\frac{\sqrt{12-3\ell^2}}{2(\ell^2+8)}, \frac{3\ell}{2(\ell^2+8)},\pm\frac{\sqrt{(\ell^2-1)(\ell^2-4)}}{2(\ell^2+8)})$\\ 
\end{itemize}
The first one is such as $y=0$ and can thus be eliminated because it does not correspond to a class 1 isotropisation. The second and the last ones do not allow to obtain $x_\pm\rightarrow 0$.  Consequently only the third equilibrium point corresponds to an isotropic state if $\ell$ tends to a constant such as $\ell^2<3$. It is similar to the one found for the flat Bianchi type $I$ model. In order to find the asymptotical behavior of $w$, we linearize (\ref{BIIeq5}) in the vicinity of equilibrium by neglecting the variables $w$ and $x_\pm$ tending to zero. It comes: 
$$ 
w\rightarrow e^{(1-\ell^2)\Omega} 
$$ 
Linearising in the same way the equations (\ref{BIIeq1}-\ref{BIIeq2}) for $x_\pm$, using the variability assumption of $\ell^2$ and introducing this last expression for $w$ , we obtain that $x_\pm$ behave as the sum of two terms $e^{2(1-\ell^2)\Omega}$ and $e^{(3-\ell^2)\Omega}$. Since isotropy needs $x_\pm\rightarrow 0$ and $\ell^2<3$, we deduce that it occurs only when $\ell^2<1$ in $\Omega\rightarrow -\infty$. The special value $\ell^2=1$ is not compatible with the isotropy. It may be deduced from our variability assumption of $\ell^2$ which implies that if $\ell^2\rightarrow  1$, $\ell^2-1$ disappears generally more quickly than $\Omega^{-1}$. But then, $w$ would tend to a non vanishing constant which is incompatible with the expression of the equilibrium points. Finally we find that asymptotically 
$$ 
x_\pm\rightarrow  e^{2(1-\ell^2)\Omega} 
$$ 
The two limits $\ell^2<1$ and $\Omega\rightarrow -\infty$ allow $x_\pm$ but also $w$ to vanish: when the Universe isotropises, it is forever expanding. In order to know if our model may isotropise, it is also necessary to check that $p_\pm e^{3\Omega}\rightarrow  0$ when $\Omega\rightarrow -\infty$. For that, we write $\dot p_\pm/H$ like a function of $x_\pm$ and $w$ and we use their asymptotic behaviors. One then calculates that $\dot p_\pm/p_\pm$ tend to the constant $-(1+\ell^2)$. Consequently, $p_\pm e^{3\Omega}\rightarrow  e^{(2-\ell^2)\Omega}$ and disappear when $\Omega$ diverges negatively and the necessary conditions for isotropy are respected. The asymptotic behaviors of the metric functions and potential are the same ones as for the Bianchi type $I$ model and depend on the same manner of disappearance or not of the function $\ell^2$ to the approach of isotropy. Concerning the 3-curvature, it tends to zero when $\Omega\rightarrow -\infty$, showing that the Universe becomes flat.
%--------------------------------------------------------------------------------------------------------------------------------------------------------------------%
\subsection{Bianchi type $VI_0$ and $VII_0$ models}\label{s2212} 
Now the variables which we will use are: 
\begin{eqnarray} 
&x_\pm=p_\pm H^{-1}&\\ 
&y=\pi R_0^3e^{-3\Omega}U^{1/2}H^{-1}&\\ 
&z=p_\phi\phi (3+2\Omega)^{-1/2}H^{-1}&\\ 
&w_\pm=\pi R_0^2e^{-2\Omega+2\beta_{+}\pm 2\sqrt{3}\beta _ -}H^{-1}\label{var41}&\\\nonumber 
\end{eqnarray} 
The difference with the Bianchi type $II$ model is that we need two variables $w_i$ to describe the curvature, one of them ($w_+$) being the variable $w$ previously defined for this last model. This is in agreement with \cite{WaiEll97} where two variables $N_i$ are also necessary for these models. The Hamiltonian constraint is written:
\begin{equation}\label{BVIcontrainte1}
x_+^2+x_-^2+24y^2+12z^2+12(w_+\pm w_-)^2=1 
\end{equation} 
and the field equations become 
\begin{eqnarray}
&\dot{x}_+=72y^2x_++24(x_+-1)(w_-\pm w_+)^2\label{BVIeq1}&\\
&\dot{x}_-=72y^2x_-+24x_-(w_-\pm w_+)^2+24\sqrt{3}(w_-^2-w_+^2)\label{BVIeq2}&\\ 
&\dot{y}=y(6\ell z+72y^2-3+24(w_-\pm w_+)^2)&\\ &\dot{z}=y^2(72z-12\ell)+24z(w_-\pm w_+)^2&\\ 
&\dot{w}_+=2w_+\left[x_++\sqrt{3}x_-+12(w_-\pm w_+)^2+36y^2-1\right]\label{BVIeq5}&\\
&\dot{w}_-=2w_-\left[x_+-\sqrt{3}x_-+12(w_-\pm w_+)^2+36y^2-1\right]\label{BVIeq6}&\\\nonumber 
\end{eqnarray} 
Anew we express the Hamiltonian equation for $H$ according to the variables $y$ and $w_\pm$:
\begin{equation}\label{BVIhamp1}
\dot{H}=-H\left[72y^2+24(w_+\pm w_-)^2\right] 
\end{equation} 
In the equations, the symbols $\pm$ correspond respectively to the models of Bianchi type $VI_0$ and $VII_0$. For the first model, the constraint (\ref{BVIcontrainte1}) shows that the variables are bounded. It is not the case for the second one: because of the minus sign, $w_+$ and $w_-$ could diverge if the difference $w_+-w_-$ remains finished, thus respecting the constraint. We will show at the end of this section that this is in fact impossible. Supposing that all the variables (but $\phi$!) are bounded allow to derive that isotropisation is impossible for a bounded value of $\Omega$. Indeed, if $\Omega\rightarrow  const$ when the proper time $t$ diverges, $d\Omega/dt\rightarrow 0$. But from the form of the lapse function and owing to the fact that $dt=-Nd\Omega$ , we deduce that $H$ should also tend to zero. It comes then from the definition of the variables $w_\pm$ that they should diverge which is incompatible with the fact that they are bounded in the vicinity of the isotropy. Thus, the isotropisation can not lead the Universe to a static state and $\Omega$ diverges inevitably.\\
We calculate the equilibrium points of the equations system (\ref{BVIeq1}-\ref{BVIeq6}). It comes: \begin{itemize} 
\item $(y, w_+, w_-)=(0, 0, 0)$ 
\item $(x_+, x_-, y, w_+, w_-)=(1, 0, 0,w_+, w_+)$ 
\item $(x_+, x_-, y, z, w_+, w_-)=(1, -\sqrt{3}, 0, 0, 0,\pm i/2)$ 
\item $(x_+, x_-, y, z, w_+, w_-)=(1, \sqrt{3}, 0, 0, \pm i/2,0)$ 
\item $(x_+, x_-, y, z, w_+, w_-)=(0, 0,\frac{\pm\sqrt{3-\ell^2}}{(6\sqrt{2})}, \ell/6, 0, 0)$ 
\item $(x_+,x_-, y, z, w_+, w_-)=(\frac{\ell^2-1}{\ell^2+8},-\sqrt{3}\frac{\ell^2-1}{\ell^2+8},
\pm\frac{\sqrt{12-3\ell^2}}{2(\ell^2+8)}, \frac{3\ell}{2(\ell^2+8)},0, \pm\frac{\sqrt{(\ell^2-1)(\ell^2-4)}}{2(\ell^2+8)})$ 
\item $(x_+,0, 0, 0, \pm i/2)$ \item $(x_+, x_-, y, z, w_+, w_-)=(1, \sqrt{3}, 0,0, \pm i/2,0)$ 
\item $(x_+, x_-, y, z, w_+, w_-)=(0, 0,\frac{\pm\sqrt{3-\ell^2}}{(6\sqrt{2})}, \ell/6, 0, 0)$ 
\item $(x_+,x_-, y, z, w_+, w_-)=(\frac{\ell^2-1}{\ell^2+8},-\sqrt{3}\frac{\ell^2-1}{\ell^2+8},\pm\frac{\sqrt{12-3\ell^2}}{2(\ell^2+8)}, \frac{3\ell}{2(\ell^2+8)},0, \pm\frac{\sqrt{(\ell^2-1)(\ell^2-4)}}{2(\ell^2+8)})$ 
\item $(x_+,x_-, y, z, w_+, w_-)=(\frac{\ell^2-1}{\ell^2+8},\sqrt{3}\frac{\ell^2-1}{\ell^2+8},\pm\frac{\sqrt{12-3\ell^2}}{2(\ell^2+8)}, \frac{3\ell}{2(\ell^2+8)},\pm\frac{\sqrt{(\ell^2-1)(\ell^2-4)}}{2(\ell^2+8)}, 0)$\\
\end{itemize} 
For the same reasons as for the Bianchi type $II$ model, the only equilibrium points corresponding to an isotropic state are: 
$$  
(x_{+}, x_{-}, y, Z, w_\pm)=(0, 0, \pm\sqrt{3-\ell^2}(6\sqrt{2})^{-1}, \ell/6, 0) 
$$ 
It will be real if $\ell^2$ tends to a constant smaller than 3.\\ 
In the same way as for the Bianchi type $II$ model, one can show that $\Omega$ is a monotonous function of the proper time whose divergence in $-\infty$ corresponds to the late times if the Hamiltonian is initially positive. We also got that the asymptotic behaviors of the functions $x_\pm$, $w_\pm$ , $p_\pm e^{3\Omega}$, $e^{-\Omega}$ and $U$ are asymptotically the same, and thus that $\ell^2<1$ when we approach the isotropy.\\ 
All these results was shown not by considering the individual behaviors of $w_+$ and $w_-$ but by considering that $w_+\pm w_-\rightarrow 0$. As we deduce from this single limit that near the isotropic state $w_\pm\rightarrow 0$, it follows that these variables are always bounded as stated at the beginning of this section, and in particular for the Bianchi type $VII_0$ model.
%--------------------------------------------------------------------------------------------------------------------------------------------------------------------%
\subsection{Bianchi type $VIII$ and $IX$ models}\label{s2213}
We will use the following variables:
\begin{eqnarray*} 
&x_\pm=p_\pm H^{-1}&\\ 
&y=\pi R_0^3e^{-3\Omega}U^{1/2}H^{-1}&\\ 
&z=p_\phi\phi (3+2\omega)^{-1/2}H^{-1}&\\ 
&w_{p}=\pi R_0^2 e^{-2\Omega+2\beta_{+}}H^{-1}&\\ &w_{m}=\pi R_0^2 e^{-2\Omega-2\beta_{+}}H^{-1}&\\
&w_{-}=e^{2\sqrt{3}\beta_{-}}&\\ 
\end{eqnarray*} 
The variables $w_p$ and $w_m$ are not independent one of the other and near an isotropic state, we have $w_p\propto w_m\propto e^{-2\Omega}H^{-1}$. Let us note moreover that $w_-$ is a positive variable. Three variables $w_i$ are thus necessary to describe the curvature in the same way that three variables $N_i$ are used in \cite{WaiEll97} for these Bianchi models. The Hamiltonian constraint is written as:
\begin{eqnarray*} 
&x_+^2+x_-^2+24y^2+12z^2+12\mbox[w_p^3(1+w_-^4)\pm 2 w_-(w_mw_p)^{3/2}(1+w_-^2)+&\\
&w_-^2(w_m^3-2w_p^3)\mbox](w_-^2w_p)^{-1}=1&\\ 
\end{eqnarray*} 
and the field equations are: 
\begin{eqnarray}
&\dot{x}_+=72y^2x_++24\{w_p^3(x_+-1)(1+w_-^4)\pm w_-(1+2x_+)(w_mw_p)^{3/2}(1+w_-^2)&\nonumber\\
&+w_-^2\left[(2+x_+)w_m^3-2(x_+-1)w_p^3\right]\}(w_-^2w_p)^{-1}\label{eq13}&\\ 
&\dot{x}_-=72y^2x_-+24\{w_p^3\left[w_-^4(x_--\sqrt{3})+x_-+\sqrt{3})\right]\pm w_-(w_mw_p)^{3/2}\mbox{[}w_-^2&\nonumber\\
&(-\sqrt{3}+2x_-)+(\sqrt{3}+2x_-)\mbox{]}+w_-^2x_-(w_m^3-2w_p^3)\}(w_-^2w_p)^{-1}\label{eq23}&\\ 
&\dot{y}=y\{6\ell z+72y^2-3+24\mbox{[}w_p^3(1+w_-^4)\pm 2(w_mw_p)^{3/2}w_-(1+w_-^2)+&\nonumber\\
&w_-^2(w_m^3-2w_p^3)\mbox{]}(w_-^2w_p)^{-1}\}&\\
&\dot{z}=y^2(72z-12\ell)+24z\mbox{[}w_p^3(1+w_-^4)\pm 2(w_mw_p)^{3/2}w_-(1+w_-^2)+&\nonumber\\
&w_-^2(w_m^3-2w_p^3)\mbox{]}(w_-^2w_p)^{-1}&\label{eq43}\\
&\dot{w}_p=w_p\{-2+2x_++72y^2+24\mbox{[}w_p^3(1+w_-^4)\pm2w_-(w_mw_p)^{3/2}(1+w_-^2)&\nonumber\\ &+w_-^2(w_m^3-2w_p^3)\mbox{]}(w_-^{2}w_p)^{-1}\}&\label{eq53}\\ 
&\dot{w}_m=w_m\{-2-2x_++72y^2+24\mbox{[}w_p^3(1+w_-^4)\pm2w_-(w_mw_p)^{3/2}(1+w_-^2)&\nonumber\\ &+w_-^2(w_m^3-2w_p^3)\mbox{]}(w_-^{2}w_p)^{-1}\}&\label{eq63}\\
&\dot{w}_-=2\sqrt{3}w_-x_-&\label{eq73}\\\nonumber 
\end{eqnarray} 
The Hamiltonian equation for $H$ becomes:
\begin{eqnarray}\label{BVIIIhamp1} 
&\dot{H}=-H\mbox{[}72y^2+24(\pm 2\frac{w_p^{1/2}w_m^{3/2}}{w_-}\pm 2w_p^{1/2}w_m^{3/2}w_--
2w_p^2+\frac{w_p^2}{w_-^2}+&\nonumber\\
&w_p^2w_-^2+\frac{w_m^3}{w_p})+\frac{3}{2}(\gamma-2)k^2\mbox{]}&\\\nonumber
\end{eqnarray} 
The signs $\pm$ respectively represent the Bianchi type $VIII$ or $IX$  models. The constraint shows that the variables are not necessarily bounded: if one of them diverges, this divergence can be compensated by that of $w_m$ or $w_p$. Thus if we show that the isotropy occurs only for  bounded values of $w_m$ and $w_p$, it will mean that it occurs only for bounded values of all the variables (but $\phi$!).\\ 
In order to achieve this goal, we will write that near an isotropic state $w_p\propto w_m
\rightarrow  w$ and $w_-\rightarrow const\not = 0$ and to simplify the explanations which will follow, we will suppose that $w_p\rightarrow w_m\rightarrow  w$ and $w_-\rightarrow 1$. Then the
constraint of the Bianchi type $VIII$ model shows that all the variables are positive and thus must take bounded values. With regard to the Bianchi type $IX$ model, let us suppose that $w$ diverges. Then if one puts $x_\pm=0$, we deduce from the constraint that $3w^2\rightarrow  2y^2+z^2-1/12$ and from the equation for $\dot w$ that $3w^2\rightarrow  3y^2-1/12$, implying that asymptotically $z^2\rightarrow y^2$ and diverge like $w^2$. However, with these limits one obtains from equations for $\dot y$ and $\dot z$ that $\dot y\rightarrow  6\ell z^2-3z$ and $\dot z\rightarrow  -12\ell z^2+2z$. Then the equilibrium for $y$ and $z$ can be only obtained when $z\rightarrow 0$ which is in contradiction with the divergence of $z$ that we have just shown. One then deduces that an isotropic equilibrium state is impossible if $w_p$ and $w_m$ diverge and that all the variables must be bounded. It follows for the same reasons as for the previous Bianchi models, that the isotropisation is impossible for a bounded value of $\Omega$.\\ 
One can also show that $w_p$ and $w_m$ can not tend to some non vanishing constants. Let us suppose that it is the case and define the two constants $w$ and $\alpha$ such as $w_p\rightarrow w$ and $w_m\rightarrow \alpha w$. We introduce these limits into the equations for $\dot x_\pm$ with $x_\pm=0$. It comes: 
\begin{eqnarray}
&\dot{x}_+=-24w^2(1+w_-\alpha^{3/2}(1+w_-^2)-2w_-^2(1+\alpha^3)+w_-^4)w_-^{-2}&\\
&\dot{x}_-=-24\sqrt{3}w^2(w_-^2-1)(1-\alpha^{3/2}w_-+w_-^2)w_-^{-2}&\\\nonumber
\end{eqnarray} 
Then, for the Bianchi type $VIII$ model, one deduces that equilibrium for $x_\pm$ will be reached only if $\alpha$ tends to the complex value $(-1)^{2/3}$ or/and if $w_-$ is negative what is impossible. For the Bianchi type $IX$ model, equilibrium for $x_\pm$ can be reached if $w_p\rightarrow w_m$ (i.e. $\beta_\pm\rightarrow  0$) and $w_-\rightarrow 1$. Then, calculating the corresponding equilibrium points, the only ones which are real and such as $w_p$ and $w_m$ are different from 0 are $(x_+, x_-, y, z, w_p, w_m, w_-)=(0, 0, \pm(6\ell)^{-1},(6\ell)^{-1}, \pm (1-\ell^2)^{1/2}(6\ell)^{-1}, 1)$. They check the constraint equation and are real if $\ell^2<1$. Moreover, one calculates that $w_p$ and $w_m$ tend to $\pm(1-\ell^2)^{1/2}(1-e^{\frac{4\Omega(\ell^2-1)+\Omega_0}{\ell^2}}+36\ell^2)^{-1/2}$ and thus reach equilibrium when $\Omega\rightarrow  +\infty$. Introducing these expressions into $\dot x_+$, it comes then that $x_+$ tends to a complex value when $\Omega\rightarrow  +\infty$ and thus that these equilibrium points are excluded.\\ 
Consequently as for the previous models, the only possible isotropic equilibrium points are such as 
$$ 
(x_{+}, x_{-}, y,z, w_p, w_m, w_-)=(0, 0, \pm \sqrt{3-\ell^2}(6\sqrt{2})^{-1}, \ell/6,0, 0, w_{-0}) $$ 
$w_{-0}$ being a constant. The variables $w_m$ and $w_p$ behave asymptotically like $e^{(1-\ell^2)\Omega}$ and $x_\pm$ like $e^{2(1-\ell^2)\Omega}$. It follows that the asymptotic behaviors of
the metric functions and the potential are the same ones asymptotically as for the other models. However, the sign of the Hamiltonian (\ref{BVIIIhamp1}) is not preserved throughout the
temporal evolution and it is thus not possible to know if the limit $\Omega\rightarrow  -\infty$ corresponds to the late or early times.
%--------------------------------------------------------------------------------------------------------------------------------------------------------------------%
\subsection{Discussion}\label{s2214} 
\begin{figure}[h ] 
\centering
\includegraphics[width=\textwidth]{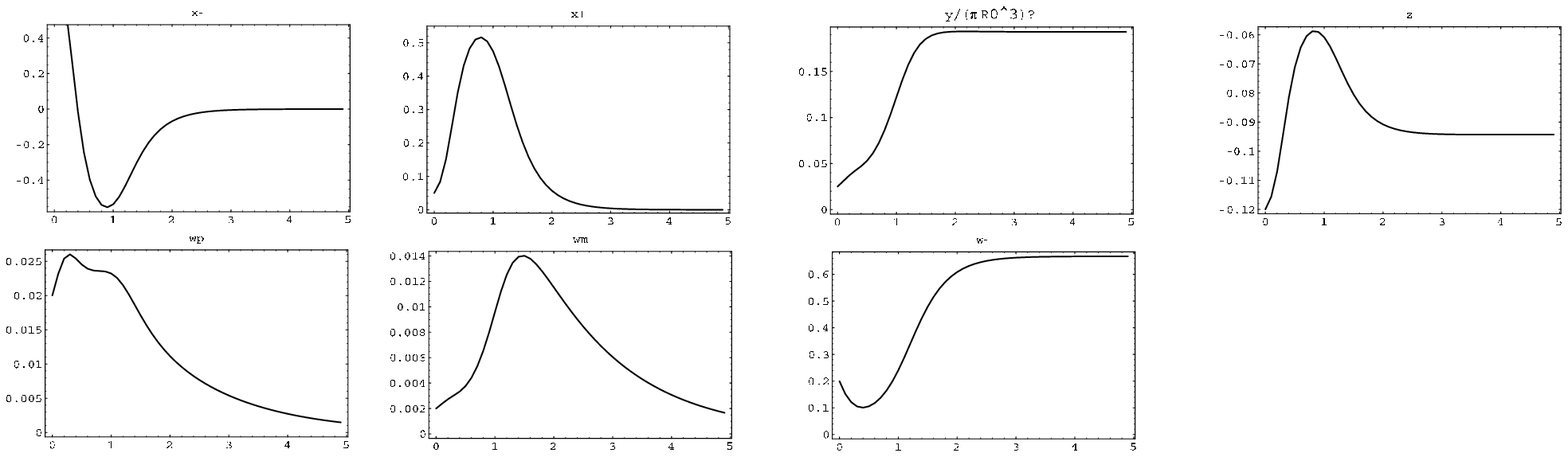}
\caption{\scriptsize{\label{BISMfig6}Evolution of the variables $x_\pm$, $y$, $z$, $w_p$, $w_m$ and $w_-$ when $\frac{(3+2\omega)^{1/2}}{\phi}=\sqrt{2}$, $U=e^{m\phi}$, $R_0^3=(\sqrt{24}\pi)^{-1}$ and $m=0.8$ with the initial values $(x_+, x_-, y, z, w_p, w_m, w_-, \phi)=(0.05, 0.83, 0.025, -0.12,0.02, 0.002, 0.2)$. All the variables tend to their theoretically predicted equilibrium values.}} 
\end{figure}

Technically, compared to the Bianchi type $I$ model, there exist several differences:
\begin{itemize} 
\item Put aside for the Bianchi type $II$ and $VI_0$ models, the constraint does not imply directly that the variables $x$, $y$, $z$ and $w$ are necessarily bounded. It must be shown that it is the case when we approach a stable isotropic state. 
\item It must be shown that the isotropy corresponds to a forever expansion of the Universe ($\Omega\rightarrow  -\infty$). 
\item It must be shown that the product $p_\pm e^{3\Omega}$ tends to zero. 
\end{itemize}
Physically, the models with curvature are more interesting than the models with flat space sections because they make it possible to show that then the isotropisation of class 1 is accompanied by an accelerated expansion and a flatness of the space sections. This comes owing to the fact that the equilibrium points are such as the variables $w$ related to the curvature disappear with the approach from isotropy, reducing the range of values in which the function $\ell$ must asymptotically tend in order to allow the isotropisation. The asymptotic behaviors of the metric functions and the potential are then the same ones as for the Bianchi type $I$ model because the Hamiltonian and the lapse function of the models with curvature behave asymptotically in the same way as for the flat model. Thus, we get the following result:
\\\\ 
\emph{Let us consider a minimally coupled and massive scalar-tensor theory and the quantity $\ell$ defined by $\ell=\frac{\phi U_\phi}{U(3+2\omega)^{1/2}}$. The asymptotic behavior of the scalar field to the approach of the isotropy is given by the asymptotic form of the solution for the differential equation $\dot\phi=2\frac{\phi^2U_\phi}{U(3+2\omega)}$ when $\Omega\rightarrow -\infty$. This limit does not necessarily correspond to the late times for the Bianchi type $VIII$ and $IX$ models contrary to the other ones. A necessary condition for a class 1 isotropisation is that $\ell^2<1$. If $\ell$ tends to a non vanishing constant, the metric functions tend to $t^{\ell^{-2}}$ and the potential disappears like $t^{-2}$.  If $\ell$ tends to vanish, the Universe tends to a De Sitter model and the potential to a constant. Whatever the curved Bianchi models, the Universe becomes flat and the expansion is accelerated.}
\\\\ 
Thus, the accelerated behavior of the Universe and its flatness could find a natural explanation
through the fact that it isotropises. Let us notice that the asymptotic behavior of the Bianchi type $IX$ model is not oscillatory. However, this is not incompatible with a mixmaster behavior in the vicinity of a singularity as noted in \cite{BarGas01}. The existence of a single isotropic equilibrium state such as the curvature tends to vanish can appear shocking but it could be due to the fact that we apply the variability assumption of $\ell$.\\ 
Considering again the scalar-tensor theory defined by $\frac{(3+2\omega)^{1/2}}{\phi}=\sqrt{2}$ and $U=e^{m\phi}$ that we studied in the previous sections, one thus finds that a necessary condition for isotropy will be $m^2<2$. A numerical integration for the case $m=0.8$ and the Bianchi type $IX$ model is illustrated on the figure \ref{BISMfig6}. 
%--------------------------------------------------------------------------------------------------------------------------------------------------------------------%
\section{Quintessence}\label{s4} 
In this section, we will show that in the vicinity of an isotropic state, the scalar field may be quintessent under certain conditions. A quintessent scalar field is such that the ratio of its pressure (\ref{pressure}) and density (\ref{density}) tends to a constant, this one having to be negative in order to be likely to explain the recent acceleration of the Universe. The scalar field pressure is then negative. To show the quintessence, one rewrites $\rho_\phi$ and $p_\phi$ using the variables of the Hamiltonian formalism. 
It comes: 
%-equation-%
\begin{equation}\label{ISQrhophi}
\rho_\phi=\frac{H^2e^{6\Omega}}{288\pi^2R_0^6}\frac{3/2+\omega}{\phi^2}\dot\phi^2+U/2
\end{equation} 
%-equation-% 
\begin{equation}\label{ISQpphi}
p_\phi=\frac{H^2e^{6\Omega}}{288\pi^2R_0^6}\frac{3/2+\omega}{\phi^2}\dot\phi^2-U/2
\end{equation} 
In what follows, we will calculate for each Bianchi models the ratio $w_\phi=p_\phi/\rho_\phi$ and will determine the conditions such that the scalar field be quintessent.
%--------------------------------------------------------------------------------------------------------------------------------------------------------------------%
\subsection{Bianchi type $I$ model without a perfect fluid}\label{ISQs111} 
Using our previous results, we find that:
$$  
H^2 e^{6\Omega}=x_0^{-2}e^{2\ell^2\Omega} 
$$ 
%-equation-%
\begin{equation}\label{ISQint}
\frac{3/2+\omega}{\phi^2}\dot\phi^2=2\ell^2 
\end{equation} 
and
$$ 
U=\frac{3-\ell^2}{72\pi^2R_0^6x_0^2}e^{2\ell^2\Omega} 
$$ 
Then , we deduce: 
$$ 
\rho_\phi=\frac{3}{144\pi^2R_0^6x_0^2}e^{2\ell^2\Omega}\propto t^{-2}
$$ 
$$ 
p_\phi=\frac{2\ell^2-3}{144\pi^2R_0^6x_0^2}e^{2\ell^2\Omega}\propto t^{-2} 
$$ 
Thus with the approach of the isotropy, the scalar field behaves like a perfect fluid with an equation of state $p_\phi=w_\phi\rho_\phi$ with $w_\phi=\frac{2}{3}\ell^2-1\in\left[-1,1\right]$ and the function of the scalar field $\ell$ can asymptotically be interpreted like the barotropic index of this fluid. It will be quintessent if $\ell^2<3/2$ what is compatible with the isotropisation. The WMAP\footnote{http://lambda.gsfc.nasa.gov/} data show that $w_\phi<-0.78$ what gives us $\ell^2<0.33$ and a deceleration parameter $q=\ell^2-1<-0.67$
%--------------------------------------------------------------------------------------------------------------------------------------------------------------------%
\subsection{Bianchi models with curvature without a perfect fluid}\label{ISQs112} 
The scalar field energy density and pressure behave in the same way but the barotropic index is this time in the range $w_\phi\in\left[-1, -1/3\right]$: the scalar field is always asymptotically quintessent when the Universe isotropises. It is in agreement with the fact that its expansion is always accelerated.
%--------------------------------------------------------------------------------------------------------------------------------------------------------------------%
\subsection{Bianchi type $I$ model with a perfect fluid such as $\Omega_m\rightarrow  0$}\label{ISQs113} 
The only thing which changes when we approach an isotropic state compared to the section \ref{ISQs111} is that $\ell^2<\frac{3}{2}\gamma$. Calculations relating to the scalar field energy density and pressure are thus the same ones as in the absence of a perfect fluid but from now on $w_\phi\in\left[-1, \gamma-1\right]$: according to the value of $\gamma$ the scalar field will be (if $\gamma\leq 1$) or not (if $\gamma\geq 1$) always quintessent.
%--------------------------------------------------------------------------------------------------------------------------------------------------------------------%
\subsection{Bianchi type $I$ model with a perfect fluid such as $\Omega_m\not\rightarrow  0$} 
In this case we have $U\propto\rho_m$ what results naturally in: 
%-equation-%
\begin{equation}\label{ISQetat3} 
U=\rho_\phi-p_\phi\propto\rho_m
\end{equation} 
If the scalar field is such as $\rho_\phi\propto p_\phi$, that means that its state equation must be the same one as that of the perfect fluid because it is the only means of explaining why their energy densities, which are obtained by writing their conservation laws, scale in the same way. Let us check it. Using the limits above, we calculate that: 
$$ 
\frac{H^2e^{6\Omega}}{144\pi^2R_0^6}\frac{3/2+\omega}{\phi^2}\dot\phi^2=\frac{\gamma^2}{64\pi^2R_0^6x_0^2\ell^2}e^{3\gamma\Omega}
$$ 
and
$$ 
U=\frac{\gamma(2-\gamma)}{64\pi^2R_0^6\ell^2x_0^2}e^{3\gamma\Omega}
$$ 
Consequently, from (\ref{ISQrhophi}-\ref{ISQpphi}), we obtain that $p_\phi=(\gamma-1)\rho_\phi$ , which is well in agreement with (\ref{ISQetat3}) and the fact that to the approach of the isotropy, the metric functions tend to $t^{\frac{2}{3\gamma}}$ as if there were only the perfect fluid and not the scalar field: obviously, the scalar field is never quintessent.
%--------------------------------------------------------------------------------------------------------------------------------------------------------------------%
\subsection{Conclusion}\label{ISQs115} 
When we consider a scalar field, we found that the constant to which tends the function $\ell$ at the time of the isotropisation can asymptotically be interpreted as being the barotropic index of the state equation characterizing the scalar field. This one can then be quintessent with
the approach of the isotropy for the Bianchi type $I$ model with or without a perfect fluid such as $\Omega_m\rightarrow  0$ if $\ell^2<3/2$. Without surprise, quintessence is impossible if $\Omega_m\not\rightarrow  0$. On the other hand, in presence of curvature it is always the
case because isotropisation imposes that $\ell^2<1$. 
%--------------------------------------------------------------------------------------------------------------------------------------------------------------------%
\section{General conclusion and perspectives}\label{s5} 
In this chapter, we considered the properties of the homogeneous cosmological models in
scalar-tensor theories with a minimally coupled and massive scalar field and studied their isotropisation process.\\ 
From the point of view of dynamical systems analysis, we detected three families of equilibrium points, corresponding to three different ways for the Universe to reach isotropy and called them class 1, 2 and 3. We were interested in class 1 and obtained results consisting of:\\ 
\begin{enumerate} 
\item the location of the stable isotropic equilibrium points 
\item some necessary conditions for their existence, constraining the scalar-tensor theories 
\item the asymptotic behaviors of the metric functions and scalar field.\\ 
\end{enumerate} 
Those of these results related to asymptotical behaviors were calculated under the assumption that the Universe tends sufficiently quickly to its equilibrium state. Mathematically speaking that means that when we approach the isotropy, the variables appearing in the field equations tend sufficiently quickly to their equilibrium values so that one can neglect their variations in the calculations. We showed how this assumption could be raised with regard to the function $\ell$. On the other hand for the other variables, a perturbative study to the approach of equilibrium could be necessary and some progress will have to be made in this direction to supplement the study of the class 1 isotropisation process.\\\\ 
Finally, the state in which the Universe is when it reaches the isotropy show some interesting characteristics which can be summarized in the following way:
\begin{itemize} 
\item the function $\ell=\phi U_\phi U^{-1} (3+2\omega)^{-1/2}$ must tend to a constant whose biggest value is constrained by the presence of curvature or perfect fluid. 
\item the Universe is expanding such as the metric functions tend to power or exponential proper time laws and the potential respectively disappears like $t^{-2}$ or tends to a constant. 
\item the presence of curvature favor late time acceleration and quintessence. 
\item the Universe is asymptotically flat. 
\end{itemize}
For the Bianchi type $I$ model, our results can be summarized in table \ref{tab2}. 
\begin{table} \begin{center}
\begin{tabular}{|c|c|c|c|} 
\hline 
&Isotropisation&Asymptotical&Quintessence\\ 
&necessary conditions&behaviors&\\
\hline 
$\Omega_m=0$&$\ell^2<3$&if $\ell^2\not\rightarrow  0$, $e^{-\Omega}\rightarrow t^{\ell^{-2}}$ &$p_\phi=(\frac{2}{3}\ell^2-1)\rho_\phi$:\\
&&and $U\rightarrow t^{-2}$&\\
&&if $\ell^2 \rightarrow  0$, $e^{-\Omega}\rightarrow e^{t}$&Yes if $\ell^2<3/2$\\ 
&&and $U\rightarrow const$&\\
\hline
$\Omega_m\rightarrow  0$&$\ell^2<\frac{3\gamma}{2}$&idem as above& idem as above\\ 
\hline 
$\Omega_m\not\rightarrow 0$&$\ell^2>\frac{3\gamma}{2}$&$e^{-\Omega}\rightarrow 
t^{\frac{2}{3\gamma}}$&$p_\phi=(\gamma-1)\rho_\phi$:\\ 
&& and $U\rightarrow t^{-2}$&No\\
\hline
\end{tabular} 
\caption{\label{tab2}\scriptsize{Isotropisation of the Bianchi type $I$ model with a minimally coupled and massive scalar field when the Universe reach sufficiently quickly its isotropic state.}} \end{center} 
\end{table} 
In presence of curvature but without a perfect fluid, the results are similar but $\ell$
must be smaller than one, thus supporting the appearance of an accelerated expansion and of quintessence.\\\\ 
Other work is in hand on this subject. Thus in \cite{a4}, for the Bianchi type $I$ model when the scalar field is not minimally coupled, i.e. when we consider a gravitation constant varying with the scalar field, we have also manage to constrain the scalar-tensor theories in such a way that they are compatible with the isotropisation. The asymptotic behaviors of the metric functions and the potential were also given but in the Einstein frame where $\phi$ is minimally coupled. It is indeed impossible to obtain these behaviors without using quadrature when the field is not minimally coupled. A case of class 2 isotropisation was detected.\\ 
In \cite{FayLum03}, always for the Bianchi type $I$ model, we generalized the results exposed in this chapter in the presence of two scalar fields. Such theories can be related to compactification process or to the presence of complex scalar fields. In this last case, we have numerically observed that isotropisation was mainly of class 3.\\ 
In \cite{a3}, we considered the Bianchi models with curvature in the presence of a perfect fluid such as $\gamma\in\left[1,2\right]$ and showed that, contrary to the Bianchi type $I$ model, the results obtained in the absence of perfect fluid were unchanged when $\Omega_m\rightarrow  0$ whereas isotropisation was impossible when $\Omega_m\not\rightarrow  0$ except if $\gamma<2/3$.\\ 
Finally in \cite{Fay03A}, we showed an interesting link between the scalar-tensor theories leading to isotropisation and those allowing a flattening of the rotation curves for the spiral galaxies: in both cases, the function $\ell$ must tend to a constant and the scalar field can be quintessent.\\\\ 
We hope having defined an operational framework  able to guide future research on the subject of the isotropisation of homogeneous but anisotropic cosmologies in scalar-tensor theories.
Amongst other things, it would be significant to take into account the cases of a negative potential or Brans-Dicke coupling function. It would then correspond to a non respect of the weak energy condition, assumption which often appears in the literature but which still misses physical motivations. More significant, it would be useful to generalize our results when the convergence to the isotropic state is not sufficiently fast to neglect the terms of the second order for the integral of $\ell^2$ (variability assumption of $\ell$) or for the bounded variables $(x_i, y, z, w_j)$ which were used to rewrite the Hamiltonian equations. Lastly, an important extension of our results would consist of the study of classes 2 and 3 isotropisation which were approached only numerically through some applications in papers that we evoke above.
\bibliographystyle{unsrt} 

\begin{thebibliography}{10}

\bibitem{Zel86}
Y~A.~B. Zel'dovich.
\newblock Cosmological field thory for observational astronomers.
\newblock {\em Sov. Sci. Rev. E Astrophys. Space Phys.}, Vol. 5:1--37, 1986.

\bibitem{Kan00}
Gordon Kane.
\newblock {\em Supersymmetry, unveiling the ultimate laws of {N}ature}.
\newblock Perseus Publishing, Cambridge, Massachusetts, 2000.

\bibitem{BraDic61}
Carl~H. Brans and Robert~H. Dicke.
\newblock Mach's principle and a relativistic theory of gravitation.
\newblock {\em Phys. Rev.}, 124, 3:925--935, 1961.

\bibitem{Bra97}
Carl~H. Brans.
\newblock Gravity and the tenacious scalar field.
\newblock {\em Contribution to Festscrift volume for Englebert Schucking},
  1997.

\bibitem{Kal21}
T.~Kaluza.
\newblock {\em Sitzungsber. Preuss. Akad. Wiss. Phys. mat. Klasse}, 966, 1921.

\bibitem{Dir37}
Paul Dirac.
\newblock {\em Nature}, 139:323, 1937.

\bibitem{Gut81}
H.~Alan Guth.
\newblock Inflationary universe: A possible solution to the horizon and
  flatness problems.
\newblock {\em Phys. Rev. D}, 23:347, 1981.

\bibitem{Car03}
Sean~M. Carroll.
\newblock {\em Why is the Universe Accelerating?}
\newblock Contribution to Measuring and Modeling the Universe, Carnegie
  Observatories Astrophysics Series Vol. 2,ed. W. L. Freedman, 2003.

\bibitem{Per99}
S.~Perlmutter et~al.
\newblock Measurements of ${\Omega}$ and ${\Lambda}$ from 42 {H}ight-{R}edshift
  {S}upernovae.
\newblock {\em Astrophysical Journal}, 517:565--586, 1999.

\bibitem{Rie98}
Adam Riess et~al.
\newblock Observational evidence from {S}upernovae for an accelerating
  {U}niverse and a cosmological constant.
\newblock {\em Astrophysical Journal}, 116:1009, 1998.

\bibitem{Bia02}
Luigi Bianchi.
\newblock {\em Rend. Accad. Naz. dei Lincei}, 11:3, 1902.

\bibitem{Lip70}
R.~Lipshitz.
\newblock {\em J. für die reine und aug. Math.}, 2:1, 1870.

\bibitem{Kil92}
W.~Killing.
\newblock {\em J. für die reine und aug. Math.}, 109:121, 1892.

\bibitem{LieEng88}
S.~Lie and F.~Engel.
\newblock {\em Theorie der Transformationsgruppen}, 1(1888) et 3(1893).

\bibitem{Tau51}
Abraham Taub.
\newblock Empty spacetimes admitting a three-parameter group of motions.
\newblock {\em Annals of Mathematics}, 53:472--490, 1951.

\bibitem{HecSch62}
O.~Heckmann and E.~Schücking.
\newblock {\em Gravitation, an Introduction to Current Research}.
\newblock Wiley, 1962.

\bibitem{EstWahBeh68}
F.B. Estabrook, W.D. Wahlquist, and C.G. Behr.
\newblock Dyadic analysis of spatially homogeneous world models.
\newblock {\em J. Math. Phys.}, 9:497--504, 1968.

\bibitem{RyaShe75}
Michael~P. Ryan and Lawrence~C. Shepley.
\newblock {\em Homogeneous Relatistic Cosmologies}.
\newblock Princeton Univ. Press, New Jersey, 1975.

\bibitem{Spe03}
D.N. Spergel et~al.
\newblock First year wilkinson microwave anisotropy probe (wmap) observations:
  Determination of cosmological parameters.
\newblock {\em Submitted to Astrophys. J.}, 2003.

\bibitem{UggElsWaiEll03}
Claes Uggla, Henk van Elst, John Wainwright, and George F.~R. Ellis.
\newblock The past atttractor in inhomogeneous cosmology.
\newblock {\em gr-qc/0304002, submitted for publication to Physical Review D},
  2003.

\bibitem{WaiEll97}
J.~Wainwright and G.F.R. Ellis, editors.
\newblock {\em Dynamical Systems in Cosmology}.
\newblock Cambridge University Press, 1997.

\bibitem{Fay01}
S.~Fay.
\newblock Isotropisation of {G}eneralised-{S}calar {T}ensor theory plus a
  massive scalar field in the {B}ianchi type {I} model.
\newblock {\em Class. Quantum Grav}, 18:2887--2894, 2001.

\bibitem{Fay01A}
S.~Fay.
\newblock Isotropisation of the minimally coupled scalar-tensor theory with a
  massive scalar field and a perfect fluid in the {B}ianchi type {I} model.
\newblock {\em Class. Quantum Grav}, 19, 2:269--278, 2002.

\bibitem{Fay03}
S.~Fay.
\newblock Isotropisation of {B}ianchi class {A} models with curvature for a
  minimally coupled scalar tensor theory.
\newblock {\em Class. Quantum Grav}, 20, 7, 2003.

\bibitem{Nar72}
Hidekazu Nariai.
\newblock Hamiltonian approach to the dynamics of {E}xpanding {H}omogeneous
  {U}niverse in the {B}rans-{D}icke cosmology.
\newblock {\em Prog. of Theo. Phys.}, 47,6:1824, 1972.

\bibitem{MatRyaTot73}
R.~A. Matzner, M.~P. Ryan, and E.~T. Toton.
\newblock The {B}rans-{D}icke theory and anisotropic cosmologies.
\newblock {\em Nuovo Cim.}, 14B:161, 1973.

\bibitem{Mis62}
C.~W. Misner.
\newblock {\em Phys. Rev.}, 125:2163, 1962.

\bibitem{ColHaw73}
C.~B. Collins and S.~W. Hawking.
\newblock Why is the universe isotropic.
\newblock {\em Astrophys. J.}, 180:317--334, 1973.

\bibitem{a4}
S.~Fay.
\newblock Isotropisation of flat homogeneous {B}ianchi type {I} model with a
  non minimally coupled and massive scalar field.
\newblock {\em To be submitted}.

\bibitem{a3}
S.~Fay.
\newblock Isotropisation of {B}ianchi class {A} models with a minimally coupled
  scalar field and a perfect fluid.
\newblock {\em Submitted}.

\bibitem{Fay00A}
S.~Fay.
\newblock Hamiltonian study of the generalized scalar-tensor theory with
  potential in a {B}ianchi type {I} model.
\newblock {\em Class. Quantum Grav.}, 17:891--902, 2000.

\bibitem{Wal83}
R.~M. Wald.
\newblock Asymptotic behavior of homogeneous cosmological models in the
  presence of a positive cosmological constant.
\newblock {\em Phys. Rev.}, D28:2118, 1983.

\bibitem{ColIbaHoo97}
A.A.Coley, J.~Ib{\`a}{\~n}ez, and R.J. van~den Hoogen.
\newblock {\em J. Math. Phys.}, 38:5256, 1997.

\bibitem{Bar97A}
John~D. Barrow.
\newblock Cosmological limits on slightly skew stresses.
\newblock {\em phys. Rev.}, D55, 12:7451, 1997.

\bibitem{SanKalWag98}
David~I. Santiago, Dimitri Kalligas, and Robert~V. Wagoner.
\newblock Scalar-{T}ensor {C}osmologies and their {L}ate {T}ime {E}volution.
\newblock {\em Phys. Rev.}, D58:124005, 1998.

\bibitem{CopLidWan98}
Edmund~J. Copeland, Andrew~R. Liddle, and David Wand.
\newblock Exponential potentials and cosmological scaling solutions.
\newblock {\em Phys. Rev.}, D57:4686--4690, 1998.

\bibitem{RosJan88}
Kjell Rosquist and Robert~T. Jantzen.
\newblock Unified regularisation of bianchi cosmology.
\newblock {\em Phys. Rep.}, 166:90--124, 1988.

\bibitem{BarGas01}
John~D. Barrow and Yves Gaspar.
\newblock {B}ianchi {VIII} empty futures.
\newblock {\em Class.Quant.Grav.}, 18:1809, 2001.

\bibitem{FayLum03}
S.~Fay and J.~P. Luminet.
\newblock Isotropisation of flat homogeneous cosmologies in presence of
  minimally coupled massive scalar fields with a perfect fluid.
\newblock {\em Submitted}, 2003.

\bibitem{Fay03A}
S.~Fay.
\newblock Scalar fields properties for flat galactic rotation curves.
\newblock {\em To be published in Astronomy and Astrophysics}, 2003.

\end{thebibliography}

\end{document}